%% file: main.tex
\documentclass[twocolumn]{article}
\usepackage{geometry}
\usepackage[compact]{titlesec} %https://tex.stackexchange.com/questions/4999/change-whitespace-above-and-below-a-section-heading
\usepackage{titling}
\usepackage{hyperref}
\hypersetup{
    colorlinks=true,
    linkcolor=blue,
    filecolor=magenta,      
    urlcolor=blue,
    citecolor=blue
    }

\usepackage{amssymb}            % Defines common symbols like \mathbb R
\usepackage{mathtools}          % Extends amsmath, providing common math tools
\usepackage{mathrsfs}           % Enables \mathscr, which can work in cases that \mathcal does not
\mathtoolsset{showonlyrefs}     % Only number equations that are referenced (optional)
\usepackage{graphicx}           % For including images
\usepackage{subcaption}         % Allows for the use of subfigures and subcaptions
\usepackage[space]{grffile}     % For spaces in image names
\usepackage{url}                % For displaying urls
\usepackage{comment}
\usepackage{tabularx}
\usepackage{ltablex}
\usepackage{tabularray}
\usepackage{multirow}
\usepackage{booktabs}
\usepackage{natbib}
\usepackage[inline]{enumitem}
\setcitestyle{numbers}
\usepackage{comment}
\usepackage{color,soul}

\usepackage{commands}
\usepackage{authblk}

\title{\textbf{Effective Monitoring of Online Decision-Making Algorithms in Digital Intervention Implementation}}

\author[1$\dagger$]{Anna L. Trella, MS}
\author[1$\dagger$]{Susobhan Ghosh, MS}
\author[2]{Erin E. Bonar, PhD}
\author[2]{Lara Coughlin, PhD}
\author[1]{Finale Doshi-Velez, PhD}
\author[3]{Yongyi Guo, PhD}
\author[4]{Pei-Yao Hung, PhD}
\author[4]{Inbal Nahum-Shani, PhD}
\author[5]{Vivek Shetty, DDS, DMD}
\author[2]{Maureen Walton, MPH, PhD}
\author[1]{Iris Yan, BS}
\author[6]{Kelly W. Zhang, PhD}
\author[1,7]{Susan A. Murphy, PhD}
\affil[1]{Department of Computer Science, Harvard University, Boston, USA}
\affil[2]{Department of Psychiatry, University of Michigan, Ann Arbor, USA}
\affil[3]{Department of Statistics, University of Wisconsin-Madison, Madison, USA}
\affil[4]{Institute for Social Research, University of Michigan, Ann Arbor, USA}
\affil[5]{School of Dentistry \& Engineering, University of California, Los Angeles, USA}
\affil[6]{Mathematics Department, Imperial College London, London, UK}
\affil[7]{Department of Statistics, Harvard University, Boston, USA}
\affil[$\dagger$]{These authors contributed equally to this work.}
\date{}
\renewcommand\Affilfont{\itshape\small}
\begin{document}
%TC:ignore
% \journaltitle{Journal Title Here}
% \DOI{DOI HERE}
% \copyrightyear{2024}
% \pubyear{-}
% \access{Advance Access Publication Date: Day Month Year}
% \appnotes{Paper}

% \firstpage{1}

% \title{Effective Monitoring of Online Decision-Making Algorithms in Digital Intervention Trials}

% \author[1$\dagger$]{Anna L. Trella, MS}
% \author[1$\dagger\ast$]{Susobhan Ghosh, MS}
% \author[2]{Erin E. Bonar, PhD}
% \author[2]{Lara Coughlin, PhD}
% \author[1]{Finale Doshi-Velez, PhD}
% \author[3]{Yongyi Guo, PhD}
% \author[4]{Pei-Yao Hung, PhD}
% \author[4]{Inbal Nahum-Shani, PhD}
% \author[5]{Vivek Shetty, PhD}
% \author[2]{Maureen Walton, PhD}
% \author[1]{Iris Yan, BS}
% \author[6]{Kelly W. Zhang, PhD}
% \author[1,7]{Susan A. Murphy, PhD}
% \affil[1]{Department of Computer Science, Harvard University, Boston, USA}
% \affil[2]{Department of Psychiatry, University of Michigan, Ann Arbor, USA}
% \affil[3]{Department of Statistics, University of Wisconsin-Madison, Madison, USA}
% \affil[4]{Institute for Social Research, University of Michigan, Ann Arbor, USA}
% \affil[5]{School of Dentistry \& Engineering, University of California, Los Angeles, USA}
% \affil[6]{Columbia University, New York, USA}
% \affil[7]{Department of Statistics, Harvard University, Boston, USA}
% % \corrauth{Abc}
% \date{}

%%%%%%%%%%%%%%%%%%%%%%%%%%%%%%%%%%%%%%%%%%%%%%%%%%%%%%%%%%%%%%%%
%% Title Page Specification
%%%%%%%%%%%%%%%%%%%%%%%%%%%%%%%%%%%%%%%%%%%%%%%%%%%%%%%%%%%%%%%%
% make title and abstract
%==============================
\begin{titlingpage}
    \maketitle
    \begin{abstract}
        \noindent\textbf{Objective: }{Online AI decision-making algorithms are increasingly used by digital interventions  to dynamically personalize treatment to individuals. These algorithms determine, in real-time, the delivery of treatment based on accruing  data.  The objective of this paper is to provide guidelines for enabling effective monitoring of  online decision-making algorithms with the goal of (1) safeguarding individuals and (2) ensuring data quality.\\} % Objective
        \textbf{Materials and Methods: }{We elucidate guidelines and discuss our experience in  monitoring online decision-making algorithms in two digital intervention clinical trials (Oralytics and MiWaves). Our guidelines include (1) developing fallback methods, pre-specified procedures executed when an issue occurs, and (2) identifying potential issues categorizing them by severity (red, yellow, and green).\\} % Materials and Methods
        \textbf{Results: }{Across both trials, the monitoring systems detected real-time issues such as out-of-memory issues, database timeout, and failed communication with an external source. Fallback methods prevented participants from not receiving any treatment during the trial and also prevented the use of incorrect data in statistical analyses.\\} % Results
        \textbf{Discussion: }{These trials provide case studies for how health scientists can build  monitoring systems for their digital intervention. Without these algorithm monitoring systems, critical issues would have gone undetected and unresolved. Instead, these monitoring systems safeguarded participants and ensured the quality of the resulting data for updating the intervention and  facilitating scientific discovery.
        % \sam{something about quality of resulting data  for data analyses and scientific discovery... } \alt{ANNA TODO: add in this.}
        \\} % Discussion
        \textbf{Conclusion: }{These monitoring guidelines and findings give digital intervention teams the confidence to include online decision-making algorithms in digital interventions.\\} % Conclusion
    \end{abstract}
\end{titlingpage}
\input{01_intro}
\input{02_case_studies}
\input{03_monitoring_guidelines}
\input{04_implementation}
\input{05_discussion}

\section*{Author Contributions}
\label{sec_author_contr}
% https://credit.niso.org/
% Contributor roles: Conceptualization, data curation, formal analysis, funding acquisition, investigation, methodology, project administration, resources, software, supervision, validation, visualization, writing - original draft, writing - review and editing.
% Conceptualization -
% Data curation -
% formal analysis - ALT, SG
% funding - 
% investigation - ALT, SG
% methodology - 
% project admin - 
% resources - VS, SAM, IN, FD, LC, EEB, MW
% software - 
% supervision - 
% validation - ALT, SG
% visualization - ALT, SG
% writing original - ALT, SG, SAM
% writing review and editing - everyone
All authors contributed to conceptualization and methodology. Data Preparation and Software (Oralytics): ALT and IY. Data Preparation and Software (MiWaves): SG and PH. Analyses: ALT and SG. Writing: ALT, SG and SAM. Funding and Administration: EEB, LC, FD, IN, VS, MW, SAM. Supervision: EEB, LC, FD, YG, IN, VS, MW, KWZ, SAM. All authors read and approved the final manuscript.

\section*{Acknowledgments}
\label{sec_ack}
This research was funded by NIH grants IUG3DE028723, P50DA054039, P41EB028242, U01CA229437, UH3DE028723, and R01MH123804.  SAM  holds concurrent appointments  at Harvard University and as an Amazon Scholar. This paper describes work performed at Harvard University and is not associated with Amazon.

%%%%%%%%%%%%%%%%%%%%%%%%%%%%%%%%%%%%%%%%%%%%%%%%%%%%%%%%%%%%%%%%
%% Bibliography
%%%%%%%%%%%%%%%%%%%%%%%%%%%%%%%%%%%%%%%%%%%%%%%%%%%%%%%%%%%%%%%%
%TC:ignore
\bibliography{main}
\bibliographystyle{unsrt}
%TC:endignore
%%%%%%%%%%%%%%%%%%%%%%%%%%%%%%%%%%%%%%%%%%%%%%%%%%%%%%%%%%%%%%%%
%% Appendices
%%%%%%%%%%%%%%%%%%%%%%%%%%%%%%%%%%%%%%%%%%%%%%%%%%%%%%%%%%%%%%%%
%TC:ignore
\input{appendix/main.tex}
%TC:endignore
\end{document}

%% file: 01_intro.tex
\section{Background and Significance}
% \section{Introduction}
\label{intro}

There is increasing interest in digital interventions that include online decision-making algorithms to personalize sequences of treatments to individuals. %\alt{Should we start with the following sentence explaining digital interventions first? The above sentence seems a bit out of place because it talks about online decision-making algorithm for personalizing treatment but we don't bring up online decision-making algorithms again until after explaining digital interventions.}
In digital interventions,  treatments involve engagement prompts containing motivational or informational messages to engage individuals in healthy behaviors.
Here personalized treatment means delivering the most beneficial treatment to an individual given their current context. Example contexts include the individual's current location, mood, 
% adherence level
and stress level.  Personalization is intended to enhance the overall effectiveness of the  digital intervention \cite{a15080255, battalio2021sense2stop,chandra2022personalization}.
% In pursuit of 
To enhance
effectiveness, online, real-time decision-making algorithms  are 
included as an intervention component in the digital intervention \cite{DBLP:journals/corr/abs-1909-03539,figueroa2021adaptive, lauffenburger2021reinforcement,forman2023using, albers2022addressing, lauffenburger2024impact, piette2022artificial}.
These algorithms  assign  treatment by identifying patterns in past  
%\alt{I noticed that in the title and abstract it seems we are moving away from clinical trials specifically and just monitoring online decision-making algorithms for digital interventions generally. If so, should we switch the language to individual and only use participant when referring to the clinical trials in the case studies?} \sam{yes, agreed...} 
data  to continuously learn and update which treatments are most effective in which contexts.

Before deploying the digital intervention, 
%in a broader health program, monitoring guidelines and 
monitoring systems must be in place to ensure safety, data quality, and implementation fidelity. These systems are well-established in the field of clinical trials.
% In the field of clinical trials, monitoring guidelines are well-established.
\begin{comment}
    Clinical trials play a critical role both in the development and evaluation of health interventions, including digital health interventions.   The field of  clinical trials has established monitoring guidelines and systems to ensure participant safety, data quality and implementation fidelity.
\end{comment}
Clinical trial monitoring traditionally involves monitoring by staff and can involve automated  monitoring by online algorithms.  Examples of monitoring systems include Data and Safety Monitoring Boards (DSMB) to oversee participant safety and data quality  \cite{national2018data}, Source Data Verification \cite{hamidi2024source} and statistical monitoring \cite{proschan2006statistical}  to ensure data quality in  clinical trials, and risk-based monitoring \cite{hurley2016risk} tools to ensure participant safety. Frameworks like FRAME \cite{wiltsey2019frame}, RE-AIM \cite{glasgow1999evaluating}, PRISM \cite{mccreight2019using} and RAPICE \cite{palinkas2019rapid} provide structured approaches for monitoring implementation fidelity \cite{breitenstein2010implementation, palmer2019understanding}, the degree at which an intervention is delivered as intended. These approaches ensure that any modifications or exceptions, along with their impact, are documented:
% and understood 
playing a crucial role in accurately interpreting findings regarding the intervention's effectiveness. 

% Monitoring implementation fidelity \cite{breitenstein2010implementation, palmer2019understanding}, the degree at which an intervention is delivered as intended, plays a crucial role in accurately interpreting findings regarding the intervention's effectiveness. 

% Frameworks like FRAME \cite{wiltsey2019frame}, RE-AIM\cite{glasgow1999evaluating}, PRISM \cite{mccreight2019using} and RAPICE \cite{palinkas2019rapid} provide structured approaches which ensure that any modifications or exceptions, along with their impact, are documented and understood.  

While many of these guidelines are useful for monitoring digital interventions, additional monitoring is required when the digital intervention includes an online decision-making algorithm.
This additional monitoring is required because the online decision-making algorithm adjusts future treatments based on an individual's past treatment responses.
% This additional monitoring is required because the online decision-making algorithm changes, as the individual experiences and responds to treatment, which treatment the individual will receive next. 
%Thus problems with the algorithm updates and its interactions with other intervention systems have to potential lead to  over/under treatment or mis-targeted treatment.
%Online decision-making algorithms require real-time updates and complex interactions with other intervention systems, all of which  impact whether or which treatment is delivered in which contexts to individuals. 
If the decision-making algorithm functions incorrectly, it detracts from the intended implementation of the digital intervention, and could compromise individuals' experience and the quality of data for use in statistical analyses \cite{trella2024monitoring}.
Minimally, such failures can 
% note for Anna:  I am not sure that we should mention, and (3)
jeopardize the future development and use of these promising
% decision-making 
algorithms.

This paper provides (1) guidelines for  monitoring  online decision-making algorithms and (2) discusses two case studies in which monitoring systems were deployed. 
These case studies offer in-depth examples and findings gained from implementing real-time monitoring systems built for digital interventions.

%% file: 02_case_studies.tex
\section{Case Studies}
\label{sec_case_studies}
% To make ideas in the rest of our paper more concrete, we use examples from two digital intervention clinical trials that have been in the field.
To illustrate the guidelines, we draw upon insights from two digital intervention clinical  trials that both employed online decision-making algorithms to personalize treatment assignment. %Each case study provides valuable contexts and considerations for developing an autonomous algorithm monitoring system.
Case study 1 describes a clinical trial of the Oralytics intervention \citep{oralytics:clinicaltrial, a15080255, trella2023reward, nahum2024optimizing, trella2024oralytics} which focuses on improving tooth-brushing behaviors for participants at risk for dental disease. Case study 2 describes a clinical trial of the MiWaves intervention \citep{miwaves:clinicaltrial, ghosh2024rebandit} that aims to reduce cannabis use amongst emerging adults (ages 18-25). In both of these digital interventions, the treatment involves the delivery of prompts to support engagement in healthy behaviors.  

%%%% FIRST MENTION OF RL %%%%
% \sam{in this section define the reward, tell reader that RL algorithm uses a model of how the reward will vary depending on the individual's current context and current treatment.} \sg{TODO: add in reward, reward model, parameters, etc. and give examples of every term defined with respect to Oralytics and MiWaves.}
The  Oralytics and MiWaves interventions each include an online decision-making algorithm to govern the delivery of  prompts or \emph{treatments}. Both decision-making algorithms use reinforcement learning (RL) \cite{sutton2018reinforcement}, a branch of machine learning focused on optimizing sequences of decisions to promote some outcome. 
% (e.g., intervention delivery decisions to promote healthy behaviors). 
% Don't really know where to put this: ``Participant data includes near time, proximal outcomes (such as OSCB) to the treatments and can include data from sensors (e.g., brushing quality, recent use of the app)as well as self-report information."
In Oralytics and MiWaves, the RL algorithm utilizes a \emph{model} of the proximal outcome or \emph{reward} (e.g., brushing quality, an individual's app engagement) that predicts how the reward will change based on the individual's current \emph{context} (e.g., time of day, recent brushing quality, recent cannabis use) 
% \sam{can we only use "state" or "context" but not both--people who are not RL experts will find confusing. }  
% \alt{Yes! I think we agreed on context so I replaced all mentions of state with context.}
and the selected \emph{treatment} or \emph{action} (e.g., delivering a  prompt vs not delivering a  prompt). 
At pre-specified \emph{update times} (e.g. nightly, weekly), the RL algorithm updates model parameters and measures of confidence in the estimates of the parameters, 
using 
% data that encompasses the 
prior interaction history (i.e., participants' context over time, the sequence of assigned treatments, and the observed rewards following each treatment assignment).  
% \sam{I edited below to try to bring in state, treatment, reward, update time, policy into the text.}
% \alt{looks good! thanks Susan!}
Using the updated model, the RL algorithm forms a \emph{policy} to assign treatment.
The RL algorithm uses this policy at \emph{decision points} (e.g., one hour before an individual brushes their teeth)
% in the morning and evening
to assign treatment according to the participant's current context.  
In this paper, we refer to the online decision-making algorithm as the RL algorithm.

\subsection{General Software Framework}
\label{sec_software_framework}
% \alt{Anna is happy with replacing sensor device with mobile app and keeping the monitoring system in the figure and just have a sentence description in the caption but not in main text for this section.}
%TC:ignore
\begin{figure*}[t]
    \centering
    % left bottom right top
    \includegraphics[width=0.8\textwidth]{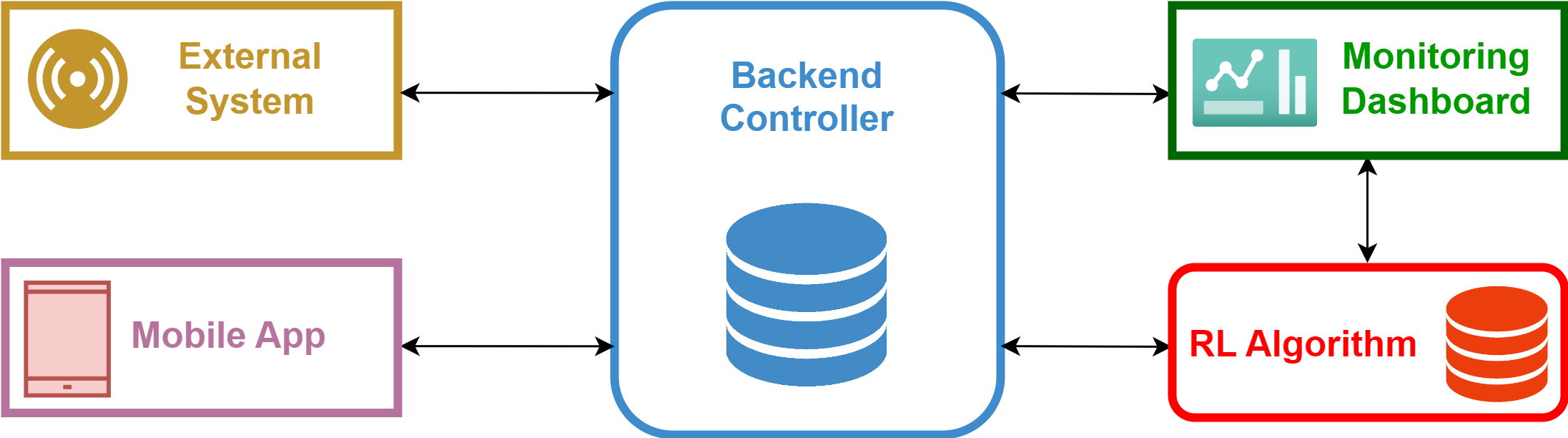}
    \caption{Overview of the software components in the intervention system for both case studies. The backend controller acts as the central coordinator and interacts with the mobile app, the RL algorithm, and external systems (i.e. third party services not controlled by the development team - like proprietary toothbrush sensors). The monitoring dashboard pulls data from the backend controller and the RL algorithm for the digital intervention staff to monitor their performance and operational status in real-time. Note that the backend controller and the RL algorithm both have their individual databases to save necessary data to reproduce decisions and facilitate subsequent statistical analyses.}
    % \sg{}
    % TODO: Caption, and change component language. Talk about sensors and monitoring dashboard}
    % \alt{1. maybe change to backend main controller?, 2. should it be Sensor Collection Devices? (e.g. Oralytics collects sensor data from both the toothbrush dock and the app)}
    % \sg{1. Backend controller is a well known term. Don't know what adding ``main'' adds here, just more confusing? 2. Don't agree with Sensor collection devices - it's either sensor devices, or sensors, or collection sensors (which is more complicated).}
    % }
    \label{fig:digital_interventions_arch}
\end{figure*}
%TC:endignore
Multiple components form the software system used in both Oralytics and MiWaves (Figure~\ref{fig:digital_interventions_arch}). 
The \emph{backend controller} (maintained by the \emph{backend controller development team}) acts as the central coordinator of the system, and manages the communication between external systems (if any), the RL algorithm (maintained by the \emph{RL development team}), and the smartphone mobile app (maintained by the \emph{app development team}). 
\emph{External systems} provide data (e.g., sensor data needed for context and reward) to the backend controller from third-party sources (e.g., a proprietary toothbrush sensor data provider). The backend controller processes this data for the \textit{RL algorithm} that assigns treatment.
For both interventions, assigning treatment means selecting between sending an engagement prompt or not. 
The backend controller obtains treatment assignments from the RL algorithm and provides them to a \emph{mobile app} that delivers engagement prompts based on the assigned treatment.  
% provides a user-friendly interface to deliver the prompts  and other intervention-related content to participants.
% The backend controller's responsibilities include:
% \begin{enumerate*}[label=(\roman*)]
%     \item requesting treatment assignments from the RL algorithm and delivering them to the mobile app,
%     \item managing data flow between various system components, such as collecting and storing sensor data,
%     \item implementing fallback mechanisms to assign treatment if the RL algorithm fails, and
%     \item monitoring for issues and raising alerts if and when issues occur.
% \end{enumerate*}
A \emph{monitoring dashboard} pulls data from all components for staff to monitor the implementation fidelity of the digital intervention and operational status in real-time. A component of this monitoring dashboard is the algorithm monitoring system (Section~\ref{alg_monitoring_sys_elems}).
% Components for the digital intervention itself (i.e., RL algorithm, mobile app, backend controller, external systems) and components supporting the digital intervention (i.e., monitoring dashboard) combine to form the software system used in both interventions (Sections \ref{sec:oralytics} and \ref{sec:miwaves}).
With the exception of the mobile app, all other components of the software system (including the RL algorithm) used in each case study were deployed on a \emph{cloud computer}.

\subsection{Case Study 1: Oralytics}
\label{sec:oralytics}
%%% talking about the intervention %%%
The Oralytics digital intervention is designed to improve the proximal outcome of oral self-care  behaviors (OSCB) for individuals 
at  risk of dental disease.  This intervention uses a commercially-available electric toothbrush with Bluetooth connectivity and integrated sensors as well as the  Oralytics mobile app. %\alt{Should we bring up the engagement prompts before mentioning the RL algorithm? I think we are switching between intervention details to RL algorithm back to intervention details and thenb ack to the RL algorithm.} \sam{I fixed} 
The intervention facilitates high-quality OSCB through self-monitoring  and prompts delivered via a push notification from the Oralytics app. These prompts contain content from the field of health psychology known to influence engagement in behavior change (reciprocity, reciprocity by proxy and curiosity) \cite{nahum2024optimizing}.
Oralytics uses an RL algorithm \cite{trella2024oralytics,trella2023reward} to govern whether or not to deliver a prompt at two decision points per day, one hour prior to an individual's usual morning and evening brushing windows. 
The proximal outcome for Oralytics is OSCB measured in seconds.
For the individual's context, the RL algorithm uses whether it was morning or evening, recent past brushing, recent number of prompts delivered, and recent app engagement (see Appendix~\ref{app_oralytics_rl_details} for exact definition of context). The RL algorithm updates its policy on a weekly cadence.
See Appendix~\ref{app_oralytics_alg_details} for more information on the  Oralytics RL algorithm.

% As discussed above, app analytics data is needed for the context and brushing data is needed in the context and proximal outcome. 
The Oralytics RL algorithm uses app analytics and brushing data. However, for the Oralytics trial, both types of data came from external systems (i.e., Oralytics app and proprietary toothbrush sensor data provider) that were developed and maintained by a separate third-party.
Thus, the RL algorithm lacked control over much of the data processing and storage.
% ; the RL algorithm relies on communication with the backend controller to obtain this data. 
We discuss later (Section~\ref{sec:context_and_constraints}) the impact of this constraint in the development of the monitoring system.

% The RL algorithm was designed to personalize when to deliver the prompts with the goal of improving the proximal outcome of brushing quality.

The  Oralytics clinical trial was a one-arm trial where all participants were provided the Oralytics digital intervention. The trial ran from September 2023 to July 2024, with $N = 79$ participants. Each participant was in the trial for 70 days. 
% The individual trial duration for participants was 70 days. The RL algorithm dynamically determined whether an intervention prompt would be delivered to  each participant twice dailyone hour before the participant's self-reported morning and evening brushing times. 
Please refer to the trial design \citep{oralytics:clinicaltrial, nahum2024optimizing} for a detailed description of the Oralytics clinical trial. 
% \sam{ Does JAMIA allow us to have cites that are not attached to a noun?   This is done repeatedly in this draft--it is quite unusual. Please email me about this!} \sg{Fixed}
% \sam{ please include cite to vivek's clinical trials.gov registration here.  Also I think sentences should end in a word, not end in a sequence of superscript numbers but you will know from looking at the jamia templates.  I leave to you guys to fix or not!}
% \alt{added vivek's clinical trials registration!}
% \sg{Susobhan, can you help with the cite numbers here as well?}

%  \sam{What is rationale for following?  Don't add unnecessary info as this gets reader off track!}, recruited incrementally with around 5 participants entering the trial every 2 weeks. 
%The trial was completed in Spring 2024.
% The trial duration for each participant was 70 days. Every day there were two decision times at which the RL algorithm decided whether to deliver an intervention prompt: (1) one hour before the participant-specified morning brushing time, and (2) one hour before the participant-specified evening brushing time. The RL algorithm updated the parameters in its model for brushing quality weekly on Sundays using all data that had been collected on all participants up to that time. See \cite{oralytics:clinicaltrial, a15080255, trella2023reward, nahum2024optimizing} for more information on the Oralytics RL algorithm and trial.

\subsection{Case Study 2: MiWaves}
\label{sec:miwaves}
% \alt{For cohesiveness with the Oralytics case study, I recommend not mentioning the 3 key intervention components in the first paragraph but directly talk about the twice daily prompts and the intervention messages and not mentioning (iii).}
%TC:ignore
% \alt{CO-AUTHOR TODOs: please take a look at this section and see if any of the ``science" of the trial or other details should be added.}
%TC:endignore
% \sam{please rewrite this case study in a parallel manner to the Oralytics case study. e.g. first overview of intervention. then a paragraph on some facts about the clinical trial.  term "individuals" in description of intervention. term "participants" in discussion of clinical trial...} 
% \alt{Oralytics section is structured as so: (1) overview of intervention, (2) description of RL algorithm and defs., and (3) clinical trial facts.} \sg{TODO}
% \sam{i tried to make parallel to other case study.} 
The MiWaves digital intervention is designed to help reduce cannabis use among emerging adults (EAs). The MiWaves intervention app includes support for self-monitoring and  prompts containing behavior change strategies.  The behavior change strategies include strategies to reduce negative and enhance positive affect, increase cannabis-free activities, and enhance future/goal-directed thinking. \cite{coughlin2024mobile}
% \sam{Susobhan please add cite to lara's in press paper here.}

%The twice daily prompts are used to facilitate an individual's self-monitoring of cannabis consumption and behavior change strategies. The individual's self-monitoring responses are also displayed as a visual feedback to promote awareness and self-reflection. On the other hand, the intervention messages focus on behavior change strategies to help individuals reduce their cannabis consumption. 

%Note that engagement with digital interventions are critical for their effectiveness. Hence, 
MiWaves uses an RL algorithm \cite{ghosh2024rebandit} to personalize, twice daily, whether or not to deliver the prompt containing a behavior change strategy.  The proximal  
outcome is  engagement with the app and the suggested behavior change strategies. The RL algorithm utilizes an individual's time-of-day, recently reported cannabis use behavior and recent app and intervention engagement as context (see Appendix \ref{app:miwaves_rl_framework} for exact definition of context).
% \alt{should we link to an appendix for the exact definition of context for MiWaves like we do for Oralytics?} 
%to personalize the likelihood of intervention message delivery, once in the morning and once in the evening. 
The RL algorithm autonomously updates its policy on a daily cadence. See Appendix \ref{app:miwaves_alg_details} for more information on the MiWaves RL algorithm.

The MiWaves RL algorithm utilizes data coming directly from the self-monitoring and mobile app usage of a participating individual, and not through an external system.
% data source maintained by a third-party - which enables the
Therefore, the RL algorithm has direct access and control over the data processing and storage. 
% This direct access ensures that the MiWaves RL algorithm can effectively utilize the data without relying on external systems.

The MiWaves clinical trial was a one-arm trial in which all participants were provided the MiWaves digital intervention. The trial ran from March 2024 to May 2024, with $N=122$ EAs. Each participant was in the trial for 30 days. 
%recruited in three waves of sizes $N=37$, $N=67$ and $N=18$ respectively.  
Please refer to the MiWaves trial design \cite{miwaves:clinicaltrial} for more information.
% \sam{separate the two cites, with one cite  with description of intervention, including RL alg., in first section and  the clinical trials cite with facts about clinical trial in the last section.} \sg{TODO}

%% file: 03_monitoring_guidelines.tex
\section{Monitoring Online Decision-Making Algorithms}

\subsection{Why Monitor?}
% \kwz{I'm wondering if this section should be split into (i) one that first reviews the components of digital health intervention trials in more detail (e.g. treatments, proximal outcomes, states/covariates, etc.), and (ii) why monitoring is needed. I think it may go a bit fast for someone in health who is unfamiliar with digital health intervention trials (depends on who is audience though).}

% \kwz{One idea is to bring in the Case study (discussion/details Oralytics and Miwaves and the description of the backend) here (so swap order of sections 2 and 3). This may illustrate more concretely all the pieces that go into a digital health trial and make the need for monitoring very concrete. 

% Then could introduce the framework for monitoring and how it is used in these two trials. This also could make the separation between the definition of concepts closer to examples of what the concepts mean, e.g., red severity concept and close by are many examples of red severity in oralytics and miwaves.}  \sam{I rather like Kelly's idea....}

%%% why pursuing online RL is challenging
Online RL algorithms generally include two processes, (1) learning: modeling how individuals' outcomes vary with treatment and updating the model's parameters (repeatedly estimated) as data accrues, and (2) treatment assignment: at each decision point, the algorithm uses the most recent  model 
%with updated parameter estimates along with 
and the individual's current context to assign treatment.  
% Ensuring that these two processes 
% % (parameter estimation and treatment assignment) 
% function correctly requires accurately collecting current context and outcome data, and sending this data to the RL algorithm to personalize treatments. 
To ensure these two processes work correctly, it’s essential to accurately collect and transmit current context and outcome data to the RL algorithm to personalize treatments.
If this data is inaccurate or missing, these two processes might cause issues that could compromise individuals' experience and the utility of data for subsequent statistical analyses. 
For example, communication issues between the RL algorithm and the backend controller can lead to individuals receiving too many or too few treatments, or treatments at the incorrect times. Or the software system 
% database 
may fail to save data  needed to conduct high-quality statistical analysis.

% In digital health trials, the online RL algorithm assigns treatment using individuals' current context and learns using data involving context, the treatment assigned for that context, and the proximal outcome occurring after treatment. Ensuring that these two processes (treatment assignment and learning) successfully function is challenging due to the unpredictable nature of the real-world environment (i.e., algorithm relies on obtaining accurate and current data to function) and the dynamic nature of online RL algorithms (i.e., treatment assignment strategy or policy is updated throughout the trial). With the promise of innovation comes the risk of any of these two processes causing issues that compromise individual experience and the utility of data for post-trial analyses.
%
% why and how monitoring helps
%Algorithm monitoring is crucial for maintaining high ethical standards in safeguarding human individuals and ensuring data quality. 
Autonomously monitoring the RL algorithm  is important for detecting, alerting, and preventing issues such as those presented above.  A monitoring system prevents such issues and detects issues as soon as they arise. This allows digital intervention teams to identify, triage, and resolve issues which (1) minimizes  negative impacts on individuals and (2) reduces the burden on the digital intervention staff who would otherwise need to manually monitor the algorithm in real time. 
% Algorithm monitoring is important for maintaining a high ethical standard of safeguarding human individuals and ensuring data quality. Autonomously monitoring an online RL algorithm during the trial is important for detecting, alerting, and preventing issues. issues and errors could arise during the trial that lead to critical situations (e.g., individuals did not get the correct treatment or database failed to save important data fields). A quality monitoring system prevents incidents and detects problems as soon as they arise. This allows research teams to immediately identify, triage, and solve issues which (1) minimizes the time and negative impact on individuals and (2) reduces the burden off of the staff in manually monitoring the trial system. 
%%%%%We now describe how to set up the autonomous monitoring system.  %%%%
% An undetected issue or an issue that takes too long to fix can significantly jeopardize trial validity, the effectiveness of the intervention, and trial results. This is especially important in the high-stakes clinical trial setting where a trial is extremely costly to run and can take years to develop and implement.

\subsection{Elements of the Algorithm Monitoring System}
\label{alg_monitoring_sys_elems}
%TC:ignore
\begin{table*}[t]
% \begin{centering}
% \begin{tabular}{p{3.5cm}p{4.2cm}p{4cm}p{4cm}}
\begin{tblr}{ Q[l,4cm] Q[c,5cm] Q[r,6.2cm]}
\hline
\hline
\textbf{Issues} & \textbf{Description} & \textbf{Example} \\
\hline
\hline
Red Severity Issues & Issues that compromise individuals' experiences or the scientific utility of data & Algorithm assigned an unreasonable number of interventions per individual over a period of time \\
\hline
Yellow Severity Issues & Issues that compromise the ability of the algorithm to learn or personalize treatment assignment & Failure to obtain current context from the backend controller \\
\hline
Green Severity Issues & Issues that need to be documented so that statistical analyses can be properly adjusted & 
% \sam{need to give an example of issue occuring with treatment assignment under green severity --otherwise this text is too vague....} 
Documentation of issue that occurred, timestamp of issue, the number of decision points and individuals impacted, how the development team fixed the issue.
\\
\hline
\hline
\textbf{Fallback Methods} & \textbf{Description} & \textbf{Example of Fallback} \\
\hline
\hline
For Treatment Assignment & Procedure executed when treatment assignment error occurs & Treatment is assigned with 0.5 probability at each affected decision point \\
\hline
For Learning & Procedure executed when algorithm update error occurs & Algorithm excludes the malformed data point when updating model parameters, and if required, postpones model update till next update time.\\
\hline
% \end{tabular}
\end{tblr}
% \end{centering}
\caption{Elements of the Algorithm Monitoring System. Potential issues are categorized into three levels of severity (red, yellow, and green) to help development teams with the order at which to resolve these issues. Fallback methods are pre-specified procedures executed to change some red severity issues into yellow severity issues and  executed when a yellow severity issue occurs. Examples are  from both Oralytics and MiWaves. 
%\sam{maybe first two rows are to prevent red/yellow severity issues???--would be good to be overt about this in the text of the document...}
%\alt{I still think that fallback methods are only triggered for yellow issues. Without these fallback methods, yellow issues would have become red issues.}
%\sam{I am ok with fallback methods making red issues into yellow issues.  The real problem is that this is all very unclear in the text.  You decide whether you want to have red issues and then explain the way you deal with red issues is via automatic deployment of fallback methods.  You can then write that this changes the red issue to a yellow issue.  The main goal is to be very clear to reader.}
%\alt{Hi Susan, Susobhan and I chatted, and the reason why we can't simply ``reduce" the severity of all red issues to yellow is because (1) sometimes the correct fallback method is unclear. Example - There is an error saving data for a particular individual in the database and (2) it takes a period of time for the check to trigger a red issue and we cannot retroactively fix. Example - You can't know what an unreasonable number of interventions over a week is until the end of that week.}
}
%\alt{We explained this point in the main text in Section 3.2 under fallback methods to be clear to the reader.}
\label{tab_elements_overview}
\end{table*}
%TC:endignore

% \kwz{This is a paper that was published in JAMIA about using adaptive learning algorithms in digital health \url{https://www.ncbi.nlm.nih.gov/pmc/articles/PMC8200266/}. Could consider making some figures and/or tables to illustrate points in the paper like they do, to enhance precision.}

We offer guidelines for building a real-time monitoring system for an RL algorithm (Table~\ref{tab_elements_overview}). These guidelines consists of: (1) issues and severity and (2) fallback methods. Both categories of guidelines work with each other to avoid issues that otherwise would have occurred. In the Section~\ref{sec_implementation}, we provide concrete examples from Oralytics and MiWaves.

\paragraph{Issues and Severity}
Development teams should begin by identifying and assigning severity to potential issues that could occur. This approach allows development teams to prioritize addressing issues based on their severity. Once these issues are identified, 
% pre-defined software packages can be utilized to send 
an automated notification system (e.g., automated email) can be utilized to alert the development teams when an issue occurs. 
% This categorization helps streamline the specification and prioritization of issues, ensuring that the most critical problems are resolved promptly.
We recommend categorizing issues into three severity levels:

\begin{itemize}
    \item \input{04_red}
    \item \input{04_yellow}
    \item \input{04_green}
\end{itemize}

\paragraph{Fallback Methods}
Fallback methods are pre-specified procedures executed when an issue occurs, serving as a backup to treatment assignment and learning procedures. These methods ensure that even when red or yellow issues occur,
% (e.g., software system issues, unavailability of sensor data for personalizing treatment, or failed communication with backend controller) 
the system defaults to baseline functionality agreed upon before deployment. 
Specifically, when treatment assignment by the RL algorithm fails, individuals should still be assigned a treatment randomized with a pre-specified probability deemed reasonable in the individual's current context. When algorithm update fails, incorrect or corrupted data should be flagged and excluded from the update.

%%% Here we talk about how fallback methods change red issues to yellow issues.
Fallback methods reduce the severity of red issues to yellow. 
%\sam{following sentence is soooooo confusing...}Therefore, fallback methods, when present, only get executed for yellow severity issues. 
However, it is not feasible to reduce all red issues to yellow because: (1) the appropriate fallback method is not always apparent (e.g., the fallback method is not apparent when there is an issue saving data needed for statistical analyses into the database)
% \sam{I don't understand this example...}); 
% \alt{discussed during Susan meeting that confusion was because not sure if this example was red or yellow issue. Made more specific that this is red because it involves data for after-study analyses NOT being saved in any database and cannot be reconstructed afterwards  and not necessarily data that affects algorithm learning.} 
and (2) the ability to identify a red issue may only occur at intervals of time, making it impossible to retroactively address them (e.g., determining whether an excessive number of interventions has occurred over a week is only possible at the end of the week).
% \sg{TODO: add in text

% \begin{itemize}
%     \item Fallback methods reduce the severity of red issues into yellow issues.
%     \item Therefore, Fallback methods only get triggered for yellow issues.
%     \item However, one cannot simply reduce the severity of all red issues to yellow issues through fallback methods because (1) sometimes the correct fallback method is unclear (e.g. there is an error saving data for a particular individual in database) and (2) it takes a period of time for the check to trigger a red issue and we canont retroactively fix (e.g. one cannot know what an unreasonable number of interventions over a week is until the end of that week).
% \end{itemize}

% }

%% file: 04_red.tex
\textbf{Red Severity (Safeguarding Individuals and Utility of Data)} Red severity issues compromise individuals' experiences or the scientific utility of the data, requiring immediate attention. These issues trigger an automated email to the algorithm development team and are prominently displayed on the monitoring dashboard for the digital intervention staff. Individuals' experiences can be compromised if they receive an unreasonable amount of prompts. This could result from communication errors between software components, 
% out-of-range
invalid treatment assignment by the RL algorithm, or in the case of a clinical trial, failing to recognize that a participant has ended the trial. The scientific utility of the data could be compromised if the database fails to store necessary values because it timed-out, closed connections, or exceeded its storage capacity.

% \textbf{Red Severity (Compromises Safeguarding Participants or Utility of Data)} Red severity issues are the most severe issues that compromise participant experience or the scientific utility of the data and require immediate attention. Red severity issues result in an immediate email to the RL software engineering team and appear on the clinician dashboard.
% Participant experience can be compromised if they receive an unreasonable amount of prompts. This could be caused by communication errors between system components (RL algorithm, main controller, app), out-of-range action-selection probabilities due to numerical instability in the RL algorithm, or failure to recognize that a participant has started in the trial. The scientific utility of the data could be compromised if the RL internal data storage fails to properly write values due to integration issues with the database management system. These issues could include failed/timed out/closed connection with the database and/or the database exceeded storage.

%% file: 04_yellow.tex
\textbf{Yellow Severity (Compromises Algorithm Functionality)}
Yellow severity issues compromise the ability of the online RL algorithm to learn and personalize treatment. 
% We recommend putting fallback methods in place to mitigate their impact.  Thus  although personalization by the RL algorithm may be compromised, the issues will  not raise to the  red severity level. 
These issues also immediately trigger an automated email and appear on the monitoring dashboard, but the urgency to resolve them is lower, allowing the RL development team more time to address them. The RL algorithm's ability to learn and personalize can be compromised in multiple ways, including (1) in the case of treatment assignment, not obtaining the necessary context data in a  timely manner, and (2) in the case of learning, failing to acquire necessary context and outcome data from the algorithm's internal database. 
% \sam{check if prior sentence is accurate!}
% \alt{it is true!}

%% file: 04_green.tex
\textbf{Green Severity (Impacts Resulting Statistical Analyses)}
Green severity issues are problems that need to be documented to properly adjust statistical analyses of the resulting data.
% These issues become problematic only if they are not documented. If not documented, these issues could compromise the validity of the statistical analyses and/or hinder the ability of the team to improve the digital intervention. 
Not documenting these issues compromises the validity of the statistical analyses and/or hinders improving the digital intervention. 
While a majority of green issues involve properly documenting red and yellow issues, one could also document issues involving components outside of the RL algorithm (e.g., mobile app or backend controller), maintain a list of items to investigate, and check the consistency of data saved by all software components.
% Managing these issues involve updating a list of items to investigate.
% Proper documentation could also involve checking the consistency of data saved by the RL algorithm's internal database with the data stored by other software components. 
Since managing these issues involve documentation by other components, we recommend establishing agreements between all software development teams prior to the deployment of the intervention. These agreements clarify responsibilities for documenting specific green issues that may arise.  
% \sam{in this section there are no examples of green severity related to issues with treatment assignment procedure.  This makes a disconnect with Table 1.}
% Green severity issues are problems that need to be documented so that post-trial analyses can be adjusted. If not properly accounted for, these issues could compromise the validity of the primary analysis or the opportunity to improve the system for future trials. Green severity issues are only problems if not documented. Handling these issues could involve a list of items to investigate after the trial is over (e.g., consistency in data saved by the main controller with data in the RL internal database). Sometimes handling these issues involves components outside of the RL system (e.g., main controller or app). Therefore, we suggest making contracts with other component teams prior to the trial to agree on who will document what data and which issues.

%% file: 04_implementation.tex
\section{Implementation of Algorithm Monitoring}
\label{sec_implementation}
\subsection{Setting and Constraints}
\label{sec:context_and_constraints}
% \alt{Differing level of trust and access and agency to other components influnece the construction of the monitoring
% system. How did the monitoring systems differ between case studies. What led them to differ. What are descriptions
% of each study that led to different systems. Broader categories depends on who is going to monitor the system.
% Trading off coverage with efficiency?}
% \sam{this paragraph will be vague for readers of JAMIA. I tried to indicate where.  Also the case studies descriptions do not provide the info necessary to really understand much of the points below. We need to mention specific constraints in an overt way.  So for example I impute from the text below that a constraint might be whether the intervention system has complete control over data processing and storage.  I think we are saying that the oralytics system was constrained due to lack of control over data processing/storage.  But this is never written explicitly.  We need to do this...  I guess another constraint was the time allocated for construction of the monitoring system.   If we are going to talk about these two constraints then we need to be really explicit about this.} 
Constraints can impact the structure of the monitoring system. 
% These constraints could involve a lack of control over data processing and storage and the limited time available to develop the monitoring system. 
For example, the constraints in Oralytics were: (1) a lack of control over data processing and storage due to dependence on an external system 
% which meant the algorithm had to coordinate with that system 
and (2) a short time frame for developing the monitoring system. In contrast, MiWaves had greater control over data processing and storage and fewer time constraints. These differing settings and constraints
% Different levels of trust,
% access, and agency between the decision-making algorithm and other trial components (e.g., sensor devices, self-report etc.) influence the type of fallback methods and the granularity and severity of issues. \sam{reader does not know what we mean by access and agency}. 
led to the development of different monitoring systems for Oralytics and MiWaves. 

Due to different time constraints, the two monitoring systems made different trade-offs between coverage of issues and efficiency in implementation. Since Oralytics had limited time to develop the monitoring system, the system specified broad categories of issues (Table~\ref{tab:oralytics_flags}). 
% Miwaves developed a more detailed and nuanced monitoring system. The MiWaves system could afford to prioritize granularity, ensuring that each issue could be identified with precision.
MiWaves had more time to develop the monitoring system and specified more detailed issues (Table~\ref{tab:miwaves_flags}). For example, while Oralytics had one broad check for failed communication with the backend controller, MiWaves had granular checks such as checking if requests from the backend controller had the correct authorization.

Since the Oralytics system only had access to sensor data (e.g., brushing data from the electronic toothbrush) from an external system after this data was pre-processed and stored in a data base maintained by the backend controller, the Oralytics monitoring system mostly focused on monitoring the validity of this data. In contrast, MiWaves only had to monitor communication issues with the backend controller because MiWaves had direct access to participant data (i.e., data was stored and processed on the MiWaves software system and not in an external system).
%a more direct data source (since the data was coming from EAs using the MiWaves app), 
% \sam{never use the word "it" or "its" in a manuscript...}
% reducing the algorithm's dependency on external endpoints and thus minimizing associated monitoring complexities. \sam{what is an "associated monitoring complexity"?} 
The differing circumstances between Oralytics and MiWaves show how different constraints 
% (e.g., control over data processing and time to develop the monitoring system) 
influence the design of monitoring systems in real-world digital intervention applications.

\subsection{Fallback Methods}
\label{fallback}
To clarify the Oralytics fallback methods presented later, we first discuss the \textit{backup treatment schedule}.
A critical Oralytics design decision was to construct a backup treatment schedule for 70 days every morning starting with the current decision point.
% rather than only the decision points for that day. 
This means treatment was selected {\it not only} for that day's decision points {\it but also} for decision points for the subsequent 69 days. See Appendix~\ref{app_backup_schedule} for full details on the backup treatment schedule. MiWaves did not employ a backup treatment schedule.

% This design decision was made because the Oralytics backend controller and mobile app were designed and developed separately by different teams (Section~\ref{sec_software_framework}). This decision mitigates possible communication issues between the backend controller and mobile app which could lead to participants failing to obtain treatment. 

% Recall that for Oralytics, the backend controller and mobile app were designed and developed separately by different teams (Section~\ref{sec_software_framework}). The decision to use a backup treatment schedule was made to mitigate possible communication issues between the backend controller and mobile app which could lead to participants failing to obtain treatment for the current decision point. On the other hand, due to the MiWaves trial design (individuals' context being always available, and the data being stored internally without depending on any external system) and greater agency between components of the software system (the MiWaves backend controller reliably requesting/obtaining treatment assignment from the RL algorithm at each decision time), the MiWaves study did not employ a backup treatment schedule. 

% To ensure that the algorithm worked consistently, 
Oralytics employed a two-layer fallback method for treatment assignment and a separate method for updating: 
\begin{enumerate*}[label=(\roman*)]
    % \item \label{oralytics_fb1} instead of just selecting a single treatment for a given decision point, the Oralytics system constructed a full treatment schedule (for the 70 days duration of the intervention) with the help of the RL algorithm for each participant in the study; and
    \item \label{oralytics_fb1} if there's any communication issue between the backend controller and the app, treatment from the most recent backup treatment schedule saved on the Oralytics app is delivered; and
    \item \label{oralytics_fb2} if there are any issues in constructing the backup treatment schedule, then treatment
    % (deliver engagement prompt versus do not deliver)
    for each decision point in the schedule is assigned with 0.5 probability rather than using the current policy. % formed from the updated model. 
    \item \label{oralytics_fb3} for updating, if issues arise (e.g.,  malformed data or recent data is not available), then the algorithm stores the data point, but does not update parameters with that data point.
\end{enumerate*}

We now describe the fallback methods for MiWaves. 
% If the RL algorithm malfunctioned (yellow severity issue), MiWaves had safeguards in place.
If the RL algorithm could not assign treatment, the MiWaves backend controller assigned treatment to participants instead (i.e. prompts were sent with $0.5$ probability). If the RL algorithm could not update its policy, the algorithm continued using the last policy to assign treatments.
% the algorithm logged the timestamps and associated causes for the update failure, and continue utilizing its old policy to assign treatments. 
This ensured that even when the algorithm update failed, the algorithm did not use malformed policies to assign treatments.

% \sam{next sentence sounds good but we are not providing evidence for this sentence.  Please provide the details here--don't reference the general section on issues in Section~\ref{sec:issues_and_severity}.  I think I get it.  What we mean is that fallback methods are deployed when ever a red or yellow severity issue occurs.  Is this correct?  } \sam{please edit this text to make this overt to reader.  Try to avoid reader struggles...} In MiWaves, the fallback mechanisms helped prevent crashes, detect issues when they occurred, and raise alerts when something went wrong (more on the issues in Section \ref{sec:issues_and_severity}).  
% \sam{sometimes we use "issues", sometimes we used "failures", and here we use "flags".  I had edited to just use "failures" but I don't have a preference.  I do want us to only use one term!}

In the  Oralytics clinical trial, all three fallback methods were triggered in response to yellow severity issues such as failing to obtain necessary data or the RL algorithm crashing. 
%(see yellow severity in Section \ref{sec:issues_and_severity} for more details). 
% \sam{I guess the following are red severity--at any rate tell reader...} 
In MiWaves, fallback methods were triggered in response to three yellow severity issues when the RL algorithm was unresponsive or had crashed. Fallback methods triggered in both trials highlight the importance of planning ahead for issues so that the digital health intervention system functions appropriately.
% whenever there are  issues either in the RL algorithm and/or in the communication between the RL algorithm and other components of the intervention software system.

\subsection{Issues and Severity}
\label{sec:issues_and_severity}
In this section, we highlight key examples and insights of issues that occurred for each severity level. For a comprehensive list of potential issues that were checked from each case study, please refer to Tables~\ref{tab:oralytics_flags} and \ref{tab:miwaves_flags} in the Appendix.
% For each issue cateogry we highlight an example of what went right and what went wrong.
% Four sets of issues that were detected/that occured during the MiWaves pilot study:
% \begin{itemize}
%     \item \textbf{Out of Memory (OOM) Errors}:
%     \item \textbf{Incorrect context feature determination}:
%     \item \textbf{Mismatch in number of active participants in the trial vs those accounted for by the RL algorithm}:
%     \item \textbf{Empty data returned by the backend controller}:
% \end{itemize}
\subsubsection{Red Severity}
Both Oralytics and MiWaves checked if 
\begin{enumerate*}[label=(\roman*)]
    \item \label{red_1} the RL algorithm assigned an unreasonable number of engagement prompts (less than the minimum amount or more than the maximum amount) and
    \item \label{red_2} treatment assignment probabilities returned by the RL algorithm for any participant at any given decision point stayed within a bounded limit (see
    % $L_{\min}, L_{\max}$ in 
    Appendix \ref{app_action_space}).
\end{enumerate*}
Due to time constraints on software development, Oralytics trial staff manually checked for red severity issues, whereas in the  MiWaves trial, this check was automated in the software and an automated email would be sent to staff and development teams if such issues were detected. 
The automation in MiWaves allowed for more comprehensive checks.
% For example, in MiWaves, the checks for \ref{red_1} included checking for an unreasonable number of assigned prompts at any decision point, and for each particular participant (measured across a span of 7 days).
For example, MiWaves included checks for an unreasonable number of assigned prompts \ref{red_1} at any decision point as well as for individual participants over seven days.
Although these types of checks were implemented, this specific category of issues was never detected in either Oralytics nor MiWaves.
% things that went bad
% \sg{please take a look to combine bottom section example}

%%% REST OF MIWAVES %%%
% \sg{Reference the bounded limit section}. 
% \alt{Can you read this and confirm if fits the definition of red severity check now?}
% \alt{yup looks good to me! just had one small comment in the text below.}
During the MiWaves trial, development teams categorized an additional set of red severity issue to detect if the cloud computer had low available memory. This precaution was taken after two incidents occurred during the trial where the available memory was too low, causing the backend controller to malfunction, and RL algorithm along with its database to crash. 
% \alt{Anna thinks the following sentence can be cleaned up and made more concise.} Notice that while the RL algorithm crashing is a yellow severity issue (due to fallback methods being present to assign treatments to participants), having low available memory on the cloud computer is a red severity issue because it has the potential to compromise the entire intervention (i.e. no interventions would be sent, impacting both participant experience and the scientific utility of data) if the backend controller also crashed due to low memory.
Low available memory on the cloud computer is a red severity issue because it can cause the backend controller to crash, preventing interventions from being sent to participants.
% \alt{maybe we should tie this back to how this issue either (i) impacts participant experinece or (ii) compromises data quality since that's why we define red issues as in earlier sections. It's hard for me to see the connection here currently.}. 
In order to preemptively mitigate the impact of this issue, the cloud computer's memory was increased, and automated email checks were designed to notify the team when the cloud computer exceeded a specified memory usage threshold. After implementation, these checks were triggered once subsequently in the MiWaves trial, which prompted the development teams to increase the available memory once again.
% \sam{too much space following this paragraph.}
% This red severity issue was triggered again when a similar issue occurred.  \sam{what do we mean by similar issue?}
% Note that until the red severity issue was addressed, the fallback method (Section~\ref{fallback}) of sending an engagement prompt with probability $0.5$ was automatically utilized.  
%     \sg{Susobhan, I am a bit confused. I feel that because the fallback method of sending an engagement prompt with probability 0.5 was executed this is not a red severity issue because the participant still gets some treatment or am I missing something?}
%     \sam{is prior sentence true?!} \sg{Anna: Let's talk about this on Monday. I think it makes sense wrt Red severity issue's definition of impacting the scientific utility of data}  \sam{I am ok with a red severity issue becoming a yellow severity due to fallback method.   I would leave the issue as red but explicitly tell reader that the issue becomes a yellow severity due to fallback method.}
% \sam{I thought we had memory issues in oralytics too and maybe Hayk forgot to tell you Anna for a while?  If so we should discuss this..} 
% \alt{We didn't have memory issues but latency issues (i.e., RL endpoints worked slower (over 5 minutes to construct schedules for everyone) the more participants entered the study). But I don't believe this caused any issues during the trial, but was a concern we brought up to fix for phase 2 because there will be more participants anticipated.}
\subsubsection{Yellow Severity} 
% Oralytics
% \sam{Throughout this section it is unclear how the RL team learned of the yellow severity issue--maybe briefly review this here? Keep parallel structure for two case studies below...}
% \alt{I added how Yellow severity issues were detected by the monitoring system and an automatic email was sent to the RL development team below.}
 
Oralytics implemented checks such as ensuring requests for sensor data (originally from external source but processed and made available to the RL algorithm by the backend controller) were successful, verifying that the sensor data was correctly formatted to be processed, and ensuring that the RL algorithm can properly read and write data to its own database. 
% Yellow severity issues were detected by the monitoring system and an automatic email was sent to the RL development team. 
% There were three types of yellow severity issues encountered during the Oralytics trial: 

One example of a yellow issue encountered during the Oralytics trial is that on two occasions, the RL algorithm crashed and was unavailable. 
However, because of fallback method~\ref{oralytics_fb1}, participants were assigned treatment from the most recent backup treatment schedule saved on their Oralytics app, but no new schedules were formed for that day. The RL development team restarted the system and the RL algorithm became available again the next day.
For additional yellow issues that occurred during the trial, see Appendix~\ref{app_oralytics_add_yellow}.

In MiWaves, the majority of the yellow severity issues were aimed at verifying the functionality of the RL algorithm and it's components, and checking for any associated issues.
For example, an yellow severity issue was encountered at the start of the MiWaves clinical trial. On the first day of the MiWaves clinical trial, certain participants were removed after the digital intervention team discovered that they were fraudulent (see Appendix \ref{app:participant_verification}). However, this information was not communicated to the RL algorithm. As a result, there was a discrepancy between the actual number of active participants in the trial and the number perceived to be active by the RL algorithm. This mismatch caused the RL algorithm to 
    %    \sam{actually I am thinking that the RL alg used data from these fraudulent participants to update the estimates of parameters in the outcome model.  This then affected treatment assignment?  Is this true?} 
    % \sg{It is true, the update used fraudulent participant data - but that did not affect treatment assignment. The bigger issue was treatment assignment getting mismatched (participant A got assigned treatment for participant B), because indexes were mismatched.} \sam{would you edit this area of text to make this clear to reader?  We need more specificity so reader does not go to sleep!} 
incorrectly assign treatments to the participants (participant A was incorrectly assigned treatment for participant B). Within one day, the development teams addressed the issue by implementing a fix (in both the backend controller and the RL algorithm) and restarting the RL algorithm to prevent similar scenarios in the future. The fix involved adding a check to communicate about active participants, and henceforth the RL algorithm maintaining the correct set of active participants in the trial. For all yellow severity issues that were encountered during the MiWaves trial, please refer to Appendix \ref{app:miwaves_yellow}.
    % \sam{give examples of specific fixes so that reader does not get bored and so reader learns something.}
% \sam{I thought the solution was that Lara's team developed a nice protocol to exclude these participants?  If this is correct, it would be cool to put this protocol online--ask Lara-- and reference it here...} \sg{Plan is to include the draft Lara has on this protocol. I have asked them for permission, so awaiting their response.}
% \end{enumerate}

\subsubsection{Green Severity}
% Oralytics
% \sam{I wonder if most of the green severity are about documenting when red/yellow flags occur or any  time a fallback method is   used?  Seems so from the descriptions below.   I suggest we discuss this as I am not quite clear in my own head about this...} \alt{I agree with you Susan. Not just the time stamps and what fallback method was used but also what fixes either the RL or backend controller software engineering team made for the issue.} \sam{great!  I started the rewrite to make this explicit. Please finish.  Also ensure discussion here and discussion in earlier section 3.2 on green severity issues is consistent.}  

Green severity issues need to be documented to adjust for statistical analyses. In both Oralytics and MiWaves, a majority of green severity incidents involved documenting red and yellow issues and information on crashes, resets, or reboots of the RL algorithm. However, information about issues with other system components (e.g., backend controller) outside of the RL algorithm were also documented. The documentation details the timestamps, the number of decision points and participants impacted, what fallback method was used for yellow issues, and how the development team fixed the issue.
During the Oralytics trial, intervention staff noticed that only a few participants had prompts scheduled on their apps. Upon investigation, the RL algorithm was functioning correctly, but the backend controller failed to wait long enough to gather the backup treatment schedules for all participants, resulting in blank schedules being sent to some participants. The issue was resolved the next business day, and the RL development team documented the incident and the affected participants and decision points.

In MiWaves, an issue occurred when the cloud computer ran out of available memory, causing the RL algorithm to crash abruptly during its daily parameter update. Since the update was not marked successful, the RL algorithm continued to use pre-crash parameters upon rebooting. The parameters were successfully updated the following day, with the timestamp recorded in the RL algorithm's database.

%% file: 05_discussion.tex
\section{Discussion}
Algorithm monitoring systems are essential to support the use of online decision-making algorithms in digital interventions. We provided guidelines for monitoring an online decision-making algorithm and shared detailed findings from two case studies where online RL algorithms dynamically adjusted treatment based on incoming data. As demonstrated in Section \ref{sec_implementation}, the absence of a monitoring system could have resulted in critical issues, such as individuals not receiving any prompts or the use of incorrect data in statistical analyses. 
% Although we showed concrete examples of monitoring systems for two RL algorithms, 
Although our paper focused on decision-making algorithms, we expect many findings are applicable to a variety of online learning algorithms, such as those used for fine-tuning prediction algorithms (e.g., predicting stress) in digital interventions. We hope these monitoring guidelines empower digital intervention teams to confidently incorporate the innovative benefits that online decision-making algorithms offer in digital intervention trials.

%% file: appendix/main.tex
\onecolumn
\appendix
\input{appendix/additional_yellow}
\input{appendix/oralytics_algorithm_details}
\input{appendix/miwaves_algorithm_details}

\input{appendix/monitoring_flags}
\input{appendix/oralytics_database_system}
\input{appendix/miwaves_database_system}

%% file: appendix/additional_yellow.tex
\section{Additional Yellow Severity Issues}
Here we list and describe all the yellow severity issues that occured during each respective trial.

\subsection{Oralytics}
\label{app_oralytics_add_yellow}
There were three types of yellow severity issues encountered during the Oralytics trial: 
\begin{enumerate}
    \item The RL algorithm was unable to obtain current context (i.e., app analytics data) from the backend controller 
    which affected the data used to update parameters. This happened on two separate occasions for two different participants: (a) 
    the first occasion was because too many requests were made as more participants had joined the trial and the backend controller communicating between the external system and the RL algorithm to provide this data  started ignoring the RL algorithm's requests after a certain threshold was exceeded and (b) the second occasion was because the raw participant's data given by the backend controller was malformed and could not be processed. In both cases, fallback method~\ref{oralytics_fb3} was executed and the RL algorithm saved the affected data points from those days for the participants affected, but did not add those data points to the batch data used for updating parameters. To solve (a), the RL development team worked with the backend controller team to create a more efficient request strategy that would not exceed the request threshold imposed by the external source and to solve (b), the RL development team worked with the backend controller team to ensure the raw participant's data was no longer malformed.
    \item On two occasions, the RL algorithm crashed and was unavailable. 
    However, because of fallback method~\ref{oralytics_fb1}, participants were  assigned treatments from the most recent backup treatment schedule saved on their Oralytics app, however, no new schedules were formed for that day. The RL development team restarted the system and the RL algorithm became available again the next day.
    \item During treatment assignment, the RL algorithm lost connection to its internal database and was unable to obtain context data from its internal database on three separate occasions. At each occasion, fallback method~\ref{oralytics_fb2} was executed and the RL algorithm provided non-personalized schedules for all affected participants. After being notified of each occasion, the RL development team restarted the RL algorithm's internal database and monitored the next week. The exact cause of this issue is still unknown and the RL team is currently investigating to prevent this issue in the next implementation.
\end{enumerate}

\subsection{MiWaves}
\label{app:miwaves_yellow}
In MiWaves, the majority of the yellow severity issues were aimed at verifying the functionality of the RL algorithm and it's components, and checking for any associated issues. Three broad instances of yellow severity issues encountered during the MiWaves clinical trial are specified below:
\begin{enumerate}
    \item \label{yellow_miwaves_scenario_1} The RL algorithm utilized consistent and accurate participant data for algorithm update. However, the RL algorithm encountered a problem prior to an update where it received duplicate or empty responses for participant data from the backend controller after the end of a given decision time.
    % \sam{hard to interpret what "each decision time" means--please edit!} 
    The development teams discovered that when a participant traveled to a different time zone and missed their self-monitoring period, the backend controller would skip the scheduled tasks for that decision time. As a result, no data was available for the RL algorithm. Since this issue occurred infrequently (in less than 3\% of decision points), the teams decided not to fix it immediately during the trial. However, the RL algorithm excluded the data associated with these decision points to update the algorithm (this was the fallback method). 
    % Moreover, the decision times, associated timestamps and affected participant information were automatically logged whenever the issue occurred. 
    %TC:ignore
    % \sam{so this is actually a green severity issue???} \sg{I added some sentences to clarify why it is yellow. Also, let me know if I should just remove the last sentence as it does talk about the green severity part of the issue, that is documenting.}
    % \sam{tell reader what was done in real time, maybe documenting xxxx with time stamps?}
    %TC:endignore

    % realized that this is due to a participant changing timezones and skipping their self check-in or the subsequent decision window end times. This skip led to the backend controller skipping the pre-scheduled tasks at the end of that particular decision time, and also not having any data to report to the RL algorithm. Since this issue occurred rarely (less than 3\% of the decision times), the trial team voted to not hotfix the issue when the trial was running.
    
    \item If the RL algorithm received an empty response concerning a particular participant from the backend controller (as seen in scenario \ref{yellow_miwaves_scenario_1}), the RL algorithm could not update the current time of day for that participant. This issue also affected all subsequent time-of-day values (necessary for the RL algorithm's context) for that participant. The app and RL development team quickly implemented a solution within two business days to fix the issue. They added a label to specify the time of day when the backend controller communicated with the RL algorithm, which the RL algorithm would then use to correctly determine the time for each participant in the trial.
    %There was an issue with the RL algorithm's determination of the time of day for a \sam{one??? or more than one??}participant. 

    % \item In relation to the previous scenario, the time of day (part of the RL algorithm's context) was being incorrectly determined after the previous decision window end call failed with an empty response from the backend controller. This meant that when the decision window end call failed, the current time of day for a participant would not get updated by the RL algorithm. This affected all subsequent time-of-day values for the participant's context. A hotfix was deployed by the app and RL development team within 2 business days that included a label to specify the time of day, which the RL algorithm would then use to determine the time of day for a given participant in the trial.
    
    \item On the first day of the MiWaves trial, certain participants were removed after the trial team discovered that they were fraudulent (see Appendix \ref{app:participant_verification}). However, this information was not communicated to the RL algorithm. As a result, there was a discrepancy between the actual number of active participants in the trial and the number perceived to be active by the RL algorithm. This mismatch caused the RL algorithm to 
    %    \sam{actually I am thinking that the RL alg used data from these fraudulent participants to update the estimates of parameters in the outcome model.  This then affected treatment assignment?  Is this true?} 
    % \sg{It is true, the update used fraudulent participant data - but that did not affect treatment assignment. The bigger issue was treatment assignment getting mismatched (participant A got assigned treatment for participant B), because indexes were mismatched.} \sam{would you edit this area of text to make this clear to reader?  We need more specificity so reader does not go to sleep!} 
    incorrectly assign treatments to the participants (participant A was incorrectly assigned treatment for participant B). Within one day, the development teams addressed the issue by implementing a fix (in both the backend controller and the RL algorithm) and restarting the RL algorithm to prevent similar scenarios in the future. The fix involved adding a flag to communicate about active participants, and henceforth the RL algorithm maintaining the correct set of active participants in the trial.
    % \sam{give examples of specific fixes so that reader does not get bored and so reader learns something.}
    % \sam{I thought the solution was that Lara's team developed a nice protocol to exclude these participants?  If this is correct, it would be cool to put this protocol online--ask Lara-- and reference it here...} \sg{Plan is to include the draft Lara has on this protocol. I have asked them for permission, so awaiting their response.}

\end{enumerate}

%% file: appendix/oralytics_algorithm_details.tex
\section{Oralytics RL Algorithm}
\label{app_oralytics_alg_details}

For completeness and to clarify ideas in the main paper, we offer an overview of key components of the Oralytics RL algorithm. See the Oralytics RL design manuscripts\cite{trella2024oralytics, nahum2024optimizing} for full details on the Oralytics RL algorithm.

\subsection{Definitions}
We use $i \in [1:N]$ to denote participants and $t \in [1:T]$ to denote decision points. $S_{i,t} \in \mathbb{R}^d$ denotes the vector of covariate features of dimension $d$ (num. of features) used for participant $i$'s current context at decision point $t$. $A_{i, t}$ denotes the treatment assigned to participant $i$'s current context at decision point $t$. Recall that in our case, assigning treatment means selecting between sending an engagement prompt or not. $Q_{i, t}$ denotes the proximal health outcome of oral self-care behaviors (OSCB) measured in seconds, observed after treatment $A_{i, t}$ is executed.

\subsection{Reinforcement Learning Framework}
\label{app_oralytics_rl_details}

\begin{itemize}
    \item \bo{Context}: Let $S_{i,t} \in \mathbb{R}^d$ represent the vector of covariate features of dimension $d = 5$ (num. of features) used as participant $i$'s current context at decision point $t$ by the RL algorithm: 
\begin{enumerate}
    \item \label{alg_context:tod} Time of Day (Morning/Evening) $\in \{0, 1\}$
    \item \label{alg_context:brushing} $\Bar{B}$: Exponential Average of Brushing Quality Over Past 7 Days (Normalized) $\in [-1, 1]$
    \item \label{alg_context:a_bar} $\Bar{A}$: Exponential Average of Prompts Sent Over Past 7 Days (Normalized) $\in [-1, 1]$
    \item \label{alg_context:app} Prior Day App Engagement (Opened App / Not Opened App) $\in \{0, 1\}$
    \item \label{alg_context:bias} Intercept Term $= 1$
\end{enumerate}

Feature 1 is 0 for morning and 1 for evening. Features 2 and 3 are $\bar{B}_{i,t} = c_{\gamma}\sum_{j = 1}^{14} \gamma^{j-1} Q_{i, t - j}$ and $\bar{A}_{i,t} = c_{\gamma}\sum_{j = 1}^{14} \gamma^{j-1} A_{i, t - j}$ respectively, where $\gamma=13/14$ and $c_{\gamma} = \frac{1 - \gamma}{1 - \gamma^{14}}$. Recall that $Q_{i, t}$ is the proximal outcome of OSCB and $A_{i,t}$ is the treatment indicator. Feature 4 is 1 if the participant has opened the app in focus (i.e., not in the background) the prior day and 0 otherwise. Feature 5 is the intercept which is always 1. Please refer to Section 2.7 in \citep{trella2024oralytics} for full details on the design of the context.

    \item \textbf{Actions}: Binary actions, i.e. $\mathcal{A} = \{0, 1\}$ - to not send an intervention message ($0$) or to send an intervention message ($1$).
    \item \textbf{Decision Points}: $T = 140$ decision points per participant. For each participant, the Oralytics clinical trial ran for $70$ days, and each day had $2$ decision points (i.e., morning and evening).
    \item \textbf{Reward}: Let $R_{i, t} \in \mathbb{R}$ denote the reward for participant $i$ after taking action $A_{i, t}$ at decision point $t$.
    We defined the reward $R_{i, t}$ given to the algorithm to be a function of brushing quality $Q_{i, t}$ (i.e., proximal health outcome) in order to improve the algorithm's learning \citep{trella2023reward}. The reward $R_{i, t}$ for the $i$th participant at decision point $t$ is:
\begin{equation}
\label{reward}
        R_{i, t} := Q_{i, t} - C_{i, t}
\end{equation}

The cost term $C_{i, t}$ allows the RL algorithm to optimize for immediate healthy brushing behavior, while also considering the delayed effects of the current action on the effectiveness of future actions. The cost term is a function (with parameters $\xi_1, \xi_2$) which takes in current context $S_{i, t}$ and action $A_{i, t}$ and outputs the delayed negative effect of currently sending a prompt.

Let $\bar{B}_{i, t}$ and $\bar{A}_{i, t}$ be context features \ref{alg_context:brushing} and \ref{alg_context:a_bar} defined earlier. The cost of sending a prompt is:
\begin{equation}
\label{cost_term}
C_{i, t} := 
\begin{cases}
\xi_1 \mathbb{I}[\bar{B}_{i, t} > b] \mathbb{I}[\bar{A}_{i, t} > a_1] & \\
\hspace{10mm} + \xi_2 \mathbb{I}[\bar{A}_{i, t} > a_2]  & \smash{\raisebox{1.6ex}{if $A_{i, t} = 1$}} \\
0 & \hspace{-0mm} \mathrm{if~} A_{i, t} = 0
\end{cases}
\end{equation}
\end{itemize}

\subsection{Treatment Assignment and Parameter Updating}
\label{app_action_space}
The algorithm assigns treatment by deciding between sending a prompt $A_{i, t} = 1$ or not $A_{i, t} = 0$. If $A_{i, t} = 1$, the backend controller (described in Section~\ref{sec_software_framework}) randomly selected content for each prompt. There are 3 different content categories: (1) participant winning a gift (direct reciprocity), (2) participant winning a gift for their favorite charity (reciprocity by proxy), and (3) Q\&A for the morning decision point or feedback on prior brushing for the evening decision point. To decide on the exact content, the main controller first samples a category with equal probability and then samples with replacement prompt content from that category. Due to the large number of prompt content in each category, it is highly unlikely that a participant received the same content twice. 
% Figure~\ref{figs/message_flow} depicts the flowchart of how a specific prompt is chosen for a patient at a decision point.

% \begin{figure}[!ht]
%     \includegraphics[width=1\textwidth]{figs/message_flow.pdf}
% \caption{{\textbf{Message Prompt Flowchart.} 
% \alt{Anna needs to make this figure prettier}
% \alt{if we are going to keep this figure, we may need to change Prompt and No Prompt to Treatment and No Treatment} There are 3 levels of randomization to select a message prompt for a patient: 
% 1. Randomization on whether to administer treatment or not (RL algorithm selects an action);
% 2. Which message category to pull from (equal probability) if the algorithm decides to administer treatment; and
% 3. Which specific prompt within the message category to use (equal probability, sampled with replacement).}}
% \label{figs/message_flow}
% \end{figure}

%%% POLICY %%%
We now discuss how the algorithm learns and assigns treatment. Since the RL algorithm in Oralytics is a generalized contextual bandit, the model of the participant environment is a model of the conditional mean reward (reward given current context and selected treatment). To model the conditional mean reward, the Oralytics RL algorithm uses a Bayesian Linear Regression with action centering model: 
\begin{equation}
\label{eqn:blr}
    R_{i, t} = S_{i, t}^T \alpha_0 + \pi_{i,t} S_{i, t}^T \alpha_1 + (A_{i, t} - \pi_{i, t}) S_{i, t}^T \beta + \epsilon_{i,t}
\end{equation}
where $S_{i, t}$ is the algorithm's context (defined in Appendix~\ref{app_oralytics_rl_details}), $\pi_{i,t}$ is the probability that the RL algorithm assigns treatment $A_{i,t} = 1$ in context $S_{i,t}$. $\epsilon_{i,t} \sim \mathcal{N}(0, \sigma^2)$ and there are priors on $\alpha_{0} \sim \mathcal{N}(\mu_{\alpha_0}, \Sigma_{\alpha_0})$, $\alpha_{1} \sim \mathcal{N}(\mu_{\beta}, \Sigma_{\beta})$, $\beta \sim \mathcal{N}(\mu_{\beta} \Sigma_{\beta})$.

\paragraph{Parameter Updating}
To learn throughout the trial, the algorithm updates its parameters. Let $\tau(i,t)$ be the update time for participant $i$ that included the current decision point $t$. We needed an additional index because update times $\tau$ and decision points $t$ are not on the same cadence. At parameter update time, the reward model's posterior parameters are updated with all history of context, treatments, and rewards up to that point. Since the RL algorithm uses Thompson sampling, the algorithm updates parameters $\mu_{\tau(i,t)}^{\text{post}}$ and $\Sigma_{\tau(i,t)}^{\text{post}}$ of the posterior distribution.

\paragraph{Treatment Assignment}
The RL algorithm for Oralytics is a modified posterior sampling algorithm called the smooth posterior sampling algorithm. The algorithm selects treatment $A_{i, t} \sim \text{Bern}(\pi_{i, t})$:

\begin{align}
    \pi_{i,t}
    = \mathbb{E}_{\tilde{\beta} \sim \mathcal{N}(\mu_{\tau(i,t) - 1}^{\text{post}}, \Sigma_{\tau(i,t) - 1}^{\text{post}} )} \left[ \rho(s^\top \tilde{\beta}) \big| \mathcal{H}_{1:n, \tau(i,t) - 1}, S_{i, t} = s \right]
\end{align}
$\tau(i,t) - 1$ is the last update time for the posterior parameters. Notice that the last expectation above is only over the draw of $\tilde{\beta}$ from the posterior distribution.
% parameterized by $\mu_{\tau(i,t) - 1}^{\text{post}}$ and $\Sigma_{\tau(i,t) - 1}^{\text{post}}$.

In smooth posterior sampling, $\rho$ is a smooth function chosen to enhance the replicability of the randomization probabilities if the study is repeated \citep{zhang2022statistical}. The Oralytics RL algorithm set $\rho$ to be a generalized logistic function:
\begin{equation}
    \rho(x) = L_{\min} + \frac{ L_{\max} - L_{\min} }{ \big[ 1 + c \exp(-b x) \big]^k}
\end{equation}
where asymptotes $L_{\min}=0.2, L_{\max}=0.8$ are clipping values (bounded away from 0 and 1) to enhance the ability to answer scientific questions with sufficient power \citep{yao2021power}. This way $\pi_{i, t}$ can only attain values within $[L_{\min}=0.2, L_{\max}=0.8]$.

% \subsection{Reward for the RL Algorithm}
% \label{app:reward}
% We defined the reward $R_{i, t}$ given to the algorithm to be a function of brushing quality $Q_{i, t}$ (i.e., proximal health outcome) in order to improve the algorithm's learning \citep{trella2023reward}. The reward $R_{i, t}$ was used to update the posterior distribution of the parameters in the reward model described in Appendix~\ref{app:action_space}. The reward for the $i$th participant at decision point $t$ was:
% \begin{equation}
% \label{reward}
%         R_{i, t} := Q_{i, t} - C_{i, t}
% \end{equation}

% The cost term $C_{i, t}$ allowed the RL algorithm to optimize for immediate healthy brushing behavior, while also considering the delayed effects of the current action on the effectiveness of future actions. The cost term is a function (with parameters $\xi_1, \xi_2$) which took in current context $S_{i, t}$ and action $A_{i, t}$ and outputted the delayed negative effect of currently sending a prompt.

% Let $\bar{B}_{i, t}$ and $\bar{A}_{i, t}$ be context features \ref{alg_context:brushing} and \ref{alg_context:a_bar} defined earlier. The cost of sending a prompt was:
% \begin{equation}
% \label{cost_term}
% C_{i, t} := 
% \begin{cases}
% \xi_1 \mathbb{I}[\bar{B}_{i, t} > b] \mathbb{I}[\bar{A}_{i, t} > a_1] & \\
% \hspace{10mm} + \xi_2 \mathbb{I}[\bar{A}_{i, t} > a_2]  & \smash{\raisebox{1.6ex}{if $A_{i, t} = 1$}} \\
% 0 & \hspace{-0mm} \mathrm{if~} A_{i, t} = 0
% \end{cases}
% \end{equation}

\subsection{Backup Treatment Schedule}
\label{app_backup_schedule}
A critical design decision made for Oralytics was to construct a backup treatment schedule (for the full 70 days duration of the intervention) every morning rather than only assigning treatment for the current day. For Oralytics, the backend controller and mobile app were designed and developed separately by different teams. Therefore, there's a higher likelihood of integration and communication issues that could arise that impact treatment delivery.
This decision mitigates possible communication issues between the backend controller and mobile app which could lead to participants failing to obtain treatment for the current decision point.

Recall that after the update of the model parameters, the RL algorithm constructs a policy based on the updated parameters. The algorithm uses the policy, combined with each participant's context to assign treatment at that decision point. For decision points of the current day, the RL algorithm has current context. However the context is unknown for future decision points. Therefore, for future decision points we use either a modified context or do not use any context and assign treatment with fixed probability. More specifically, the backup treatment schedule is constructed as follows:
\begin{itemize}
    \item For the current day's decision points, $j = t, t + 1$, the algorithm uses the current context $S_{i, t}, S_{i, t + 1}$ and the current policy to assign treatment.
    \item For decision points within the first 2 weeks from $t$, $j = t + 2, t + 3,..., t + 26, t + 27$, the algorithm uses a modified context (see Appendix~\ref{app:modified_rl_features}) and the current policy to assign treatment.
    \item For all decision points after $t + 27$, the algorithm assigns treatment with a fixed probability $0.5$.
\end{itemize}

\subsubsection{Modified Feature Space for Fallback Method}
\label{app:modified_rl_features}
For decision points within the first 2 weeks from the current day, the algorithm uses the context specified in Appendix~\ref{app_oralytics_rl_details} with the following modifications:
\begin{itemize}
    \item Feature~\ref{alg_context:brushing} is replaced with the \textit{Most Recent} Exponential Average Brushing Quality Over Past 7 Days (Normalized)
    \item Feature~\ref{alg_context:app} is replaced with the \textit{Best Guess of} Prior Day App Engagement (Opened App / Not Opened App) $= 0$
\end{itemize}

Of the context features presented earlier, features \ref{alg_context:bias}, \ref{alg_context:tod}, and \ref{alg_context:a_bar} are known except for features \ref{alg_context:brushing} ($\bar{B}$) and \ref{alg_context:app} (prior day app engagement). Since future values of $\bar{B}$ are unknown, we impute this feature with the most recent value of $\bar{B}_{i, t}$ known when the schedule was formed. Namely, the $\bar{B}$ value for decision points $j = t + 2, ...,t + 27$ used  the same $\bar{B}$ value as decision points $t, t + 1$.  For the prior day app engagement feature, we impute the value to 0, because our best guess is that the participant does not getting a fresh schedule because they did not open the app.

%% file: appendix/miwaves_algorithm_details.tex
\section{MiWaves RL Algorithm}
\label{app:miwaves_alg_details}
For completeness and to clarify ideas in the main paper, we also offer an overview of key components of the MiWaves RL algorithm. Please refer to the MiWaves RL algorithm design manuscripts\cite{coughlin2024mobile, ghosh2024rebandit} for full details on the MiWaves RL algorithm. We use $i \in [1:m]$ to denote participants and $t \in [1:T]$ to denote decision points.

\subsection{Reinforcement Learning Framework}
\label{app:miwaves_rl_framework}
This section provides a brief overview of the Reinforcement Learning (RL) framework used in this work with respect to the MiWaves trial.

\begin{itemize}
    \item \bo{Context}: Let $\state{t}{i} \in \mathbb{Z}^d_2$ represent the vector of binary covariates of dimension $d = 3$ (number of features) used as participant $i$'s current context at a given decision point $t$ by the RL algorithm. It is defined as a 3 dimensional tuple of binary features, i.e. $\state{t}{i} = \{S_{i1}^{(t)}, S_{i2}^{(t)}, S_{i3}^{(t)}\}$ where:
    \begin{itemize}
        \item $S_{i1}^{(t)}$: Recent app engagement (Low / High) over the past 3 days $\in \{0, 1\}$
        \item $S_{i2}^{(t)}$: Time of day (Morning / Evening) $\in \{0, 1\}$
        \item $S_{i3}^{(t)}$: Recent cannabis use reported during the most recent self-monitoring period (Using / Not using) $\in \{0, 1\}$
    \end{itemize}
    $S_{i2}^{(t)}$ is 0 for morning and 1 for evening. For $S_{i1}^{(t)}$ and $S_{i3}^{(t)}$, the less favorable states are represented as 0 (using cannabis, low app engagement), and the more favorable ones are represented as 1 (not using cannabis, high app engagement). 
    % Please refer to the MiWaves RL design manuscript\citep{miwaves_rl} for full details on the design of the state space.
    Please refer to Section 1.3 in the MiWaves RL design manuscript\citep{miwaves_rl} for full details on the design of the state space.

    \item \textbf{Actions}: Binary actions, i.e. $\mathcal{A} = \{0, 1\}$ - to not send an engagement prompt ($0$) or to send an engagement prompt ($1$).
    \item \textbf{Decision Points}: $T = 60$ decision points per participant. For each participant, the MiWaves clinical trial ran for $30$ days, and each day had $2$ decision points per day. Therefore, we had $60$ decision points per participant.
    \item \textbf{Reward}: Let us denote the reward for participant $i$ after taking an action at decision point $t$ by $\reward{t+1}{i} \in \{0, 1, 2, 3\}$. One of the assumptions underpinning the MiWaves clinical trial is that higher app and intervention engagement (self-monitoring activities and using suggestions by MiWaves to manage mood, identify alternative activities, plan goals, etc.)  will lead to lower cannabis use. Hence, the higher levels of reward correspond to higher levels of engagement by the participant, and vice-versa.
\end{itemize}

\subsection{Treatment Assignment and Parameter Updating}
\label{app:miwaves_model}

%%% POLICY %%%
We now discuss how the algorithm learns and assigns treatment. Since the RL algorithm in MiWaves is a generalized contextual bandit, the model of the participant environment is a model of the conditional mean reward (reward given current context and selected treatment). To model the conditional mean reward, the MiWaves RL algorithm utilizes a Bayesian Linear Mixed Model with action centering \cite{greenewald2017action}. For a given participant $i$ at decision point $t$, the RL algorithm receives the reward $\reward{t+1}{i}$ after taking action $\action{t}{i}$ in the participant's current state $\state{t}{i}$. Then, the algorithm's reward approximation model for participant $i$ is written as: 
\begin{align}
    \reward{t+1}{i}
    &= g(\state{t}{i})^T \bs{\alpha_i}  + (\action{t}{i} - \pii{t}{i}) f(\state{t}{i})^T \bs{\beta_i} + (\pi_i^{(t)})f(\state{t}{i})^T \bs{\gamma_i} + \epsilon_i^{(t)}
\end{align}
where $\epsilon_i^{(t)}$ is the noise, assumed to be gaussian i.e. $\bs{\epsilon} \sim \mathcal{N}(\bs{0}, \sigma_{\epsilon}^2\bs{I}_{t m_t})$, and $m_t$ is the total number of participants who have been or are currently part of the MiWaves digital intervention at time $t$. Also $\bs{\alpha_i}$, $\bs{\beta_i}$, and $\bs{\gamma_i}$ are weights that the algorithm wants to learn. $\pii{t}{i}$ is the probability of taking action $\action{t}{i} = 1$ in state $\state{t}{i}$ for participant $i$ at decision point $t$. $g(\state{t}{i})$ and $f(\state{t}{i})$ are functions of the algorithm's context (defined in Appendix \ref{app:miwaves_rl_framework}), defined as:
\begin{align}
    g(\state{t}{i}) = f(\state{t}{i}) = [1, S^{(t)}_{i1}, S^{(t)}_{i2}, S^{(t)}_{i3}, S^{(t)}_{i1} S^{(t)}_{i2}, S^{(t)}_{i2} S^{(t)}_{i3}, S^{(t)}_{i1} S^{(t)}_{i3}, S^{(t)}_{i1} S^{(t)}_{i2} S^{(t)}_{i3}]
\end{align}

We refer to the term $g(\state{t}{i})^T \bs{\alpha_i}$ as the baseline, and $f(\state{t}{i})^T \bs{\beta_i}$ as the advantage (i.e. the advantage of taking action 1 over action 0). 

We re-write the reward model as follows:
\begin{align}
    \reward{t+1}{i} 
    &= \Phii{T}{it} \tparam{}{i} + \epsilon_{i}^{(t)}
\end{align}
where $\Phii{T}{it} = \Phii{}{}(\state{t}{i}, \action{t}{i}, \pii{t}{i})^T = [g(\state{t}{i})^T, (\action{t}{i} - \pii{t}{i}) f(\state{t}{i})^T, (\pi_i^{(t)})f(\state{t}{i})^T]$ is the design matrix for given state and action, and $\tparam{}{i} = [\bs{\alpha_i}, \bs{\beta_i}, \bs{\gamma_i}]^T$ is the joint weight vector that the algorithm wants to learn. We further break down the joint weight vector $\tparam{}{i}$ into two components:
\begin{align}
    \tparam{}{i} = \begin{bmatrix}
            \bs{\alpha}_i\\
            \bs{\beta}_i\\
            \bs{\gamma}_i
        \end{bmatrix}
        = \begin{bmatrix}
        \bs{\alpha}_{\text{pop}} + \bs{u}_{\alpha, i}\\
        \bs{\beta}_{\text{pop}} + \bs{u}_{\beta, i}\\
        \bs{\gamma}_{\text{pop}} + \bs{u}_{\gamma, i}
        \end{bmatrix}
        = \tpop{} + \ui{}{i}
\end{align}
Here, $\tpop{}{} = [\bs{\alpha}_{\text{pop}}, \bs{\beta}_{\text{pop}}, \bs{\gamma}_{\text{pop}}]^T$ is the population level term which is common across all the participant's reward models. We assume a gaussian prior distribution on $\tpop{}$ given by $\tpop{} \sim \mathcal{N}(\muprior, \Sigprior)$. On the other hand, $\ui{}{i} = [\bs{u}_{\alpha, i}, \bs{u}_{\beta, i}, \bs{u}_{\gamma, i}]^T$ are the individual level parameters, or the \emph{random effects}, for any given participant $i$. Note that the individual level parameters are assumed to be gaussian by definition, i.e. $\ui{}{i} \sim \mathcal{N}(\bs{0}, \Sig{}{u})$. 

\paragraph{Parameter Updating}
To learn, the algorithm updates its parameters daily (after every even numbered decision point $t$). At parameter update time, the reward model's posterior parameters are updated with all history of context, treatments, and rewards up to that point. The MiWaves RL algorithm uses a variant of Thompson sampling, and hence the algorithm updates parameters $\bs{\mu^{(t)}_{\text{post}}}$ and $\Sig{(t)}{\text{post}}$ of the posterior distribution. The updated posteriors are given as:

\allowdisplaybreaks
\begin{align}
    \bs{\mu^{(t)}_{\text{post}}} &= \bigg(\bs{\Tilde{\Sigma}^{-1}_{\theta, t}} + \frac{1}{\sige{2}}\bs{A} \bigg)^{-1} \bigg( \bs{\Tilde{\Sigma}^{-1}_{\theta, t}} \bs{\mu_{\theta}} + \frac{1}{\sige{2}}\bs{B} \bigg) \label{eqn:postMeanTheta} \\
    \Sig{(t)}{\text{post}} &= \bigg(\bs{\Tilde{\Sigma}^{-1}_{\theta, t}} + \frac{1}{\sige{2}} \bs{A} \bigg)^{-1} \label{eqn:postCovTheta}
\end{align}
where
\begin{align}
    \bs{A} &=
    \begin{bmatrix}
        \sum_{\tau=1}^t \Phii{}{1\tau} \Phii{T}{1\tau} & \bs{0} &\cdots &\bs{0}\\
        \bs{0} & \sum_{\tau=1}^t \Phii{}{2\tau} \Phii{T}{2\tau} & \cdots & \bs{0} \\
        \vdots & \vdots & \ddots & \vdots\\
        \bs{0} & \bs{0} & \cdots & \sum_{\tau=1}^t \Phii{}{m_t\tau} \Phii{T}{m_t\tau}
    \end{bmatrix}\\
    \bs{B} &=
    \begin{bmatrix}
        \sum_{\tau=1}^t \Phii{}{1\tau} R^{(\tau + 1)}_{1}\\
        \vdots\\
        \sum_{\tau=1}^t \Phii{}{m_t\tau} R^{(\tau + 1)}_{m_t}\\
    \end{bmatrix} \; \; \;
    \bs{\mu_{\theta}} =
    \begin{bmatrix}
        \bs{\muprior} \\
        \bs{\muprior} \\
        \vdots\\
        \bs{\muprior}
    \end{bmatrix} \label{eqn:mu_t0}\\
    \bs{\Tilde{\Sigma}_{\theta, t}} &= 
    \begin{bmatrix}
        \bs{\Sigprior} + \bs{\Sigma_{u}} & \bs{\Sigprior} & \cdots & \bs{\Sigprior}\\
        \bs{\Sigprior} & \bs{\Sigprior} + \bs{\Sigma_{u}} & \cdots & \bs{\Sigprior}\\
        \vdots & \vdots & \ddots & \vdots\\
        \bs{\Sigprior} & \bs{\Sigprior} & \cdots & \bs{\Sigprior} + \bs{\Sigma_{u}}
    \end{bmatrix} \label{eqn:sig_tt}
\end{align}

The estimates of $\sige{2}$ and $\bs{\Sigma_{u}}$ are updated at the end of each week. The update procedure for these estimates can be found in Section 2.2.2 of the MiWaves RL design manuscript \cite{miwaves_rl, ghosh2024rebandit}.

\paragraph{Treatment Assignment}
The MiWaves RL algorithm utilizes a variant of posterior sampling called the smooth posterior sampling for treatment assignment. The treatment assignment procedure utilizes the Gaussian posterior distribution defined by the posterior mean $\bs{\mu^{(t-1)}_{\text{post}}}$ and variance $\Sig{(t-1)}{\text{post}}$ to determine the action selection probability $\pii{t}{}$ and the corresponding action $\action{t}{}$ for the given time step $t$ as follows:

\begin{align}
    \pi_i^{(t)} &= \mathbb{E}_{{\Tilde{\beta} \sim \mathcal{N}(\mu_{\text{post}, i}^{(t-1)}, \Sigma_{\text{post}, i}^{(t-1)})}}[\rho(f(\state{t}{i})^T \boldsymbol{\Tilde{\beta}}) |\mathcal{H}_{1:m_{t-1}}^{(t-1)},  \state{t}{i}]
\end{align}
where $\mathcal{H}^{(t-1)}_{1:m_{t-1}} = \{\state{1}{i}, \action{1}{i}, \reward{2}{i}, \cdots, \state{t-1}{i}, \action{t-1}{i}, \reward{t}{i}\}_{i \in [m_{t-1}]}$ refers to the trajectories (history of state action reward tuples) from time $\tau = 1$ to $\tau = t-1$ for all participants $i \in [m_{t-1}]$.
Notice that the last expectation above is only over the draw of $\beta$ from the posterior distribution parameterized by $\boldsymbol{\mu_{\text{post}, i}^{(t-1)}}$ and $\boldsymbol{\Sigma_{\text{post}, i}^{(t-1)}}$.

In smooth posterior sampling, $\rho$ is a smooth function chosen to enhance the replicability of the randomization probabilities if the trial is repeated \citep{zhang2022statistical}. The MiWaves RL algorithm set $\rho$ to be a generalized logistic function:
\begin{equation}
    \rho(x) = L_{\min} + \frac{ L_{\max} - L_{\min} }{ \big[ 1 + c \exp(-b x) \big]^k}
\end{equation}
where asymptotes $L_{\min}=0.2, L_{\max}=0.8$ are clipping values (bounded away from 0 and 1) to enhance the ability to answer scientific questions with sufficient power \citep{yao2021power}. This way $\pi_i^{(t)}$ can only attain values within $[L_{\min}=0.2, L_{\max}=0.8]$. The algorithm assigns treatment by deciding between sending a prompt $a_{i}^{(t)} = 1$ or not $a_{i}^{(t)} = 0$ as follows:
\begin{align}
    \action{t}{i} &\sim \text{Bern}(\pii{t}{i})
\end{align}

If $a_{i}^{(t)} = 1$, the backend controller (described in Section~\ref{sec_software_framework}) randomly selects the intervention content for each prompt. In the MiWaves clinical trial, there were 6 different content categories depending on the prompt length and the requested interaction level of the participant. More details on the content categories can be found in Table 2 of the MiWaves protocol\cite{coughlin2024mobile}.

%% file: appendix/monitoring_flags.tex
\section{Potential Issues for Oralytics}

% \begin{table*}[t]
%     \centering
    \begin{tabularx}{\textwidth}{ |c|c|X| }
    \hline
    Severity & EC & Description \\
    \hline
    \hline
    \multirow{5.5}{*}{RED} & R1 & Algorithm has assigned no treatment to participant $\{\text{\texttt{participant id}}\}$ for every decision point in the last 7 days \\
    \cline{2-3}
    & R2 & Algorithm has assigned treatment to participant $\{\text{\texttt{participant id}}\}$ for every decision point in the last 7 days \\
    \cline{2-3}
    & R3 & Error in saving data for participant $\{\text{\texttt{participant id}}\}$. \\
    \hline
    \hline
    \multirow{9}{*}{YELLOW} & Y1 & Dependency endpoint call to $\{\text{\texttt{endpoint name}}\}$ failed for user $\{\text{\texttt{participant id}}\}$. \\
    \cline{2-3}
    & Y2	& Endpoint response to $\{\text{\texttt{endpoint name}}\}$ cannot be json-ified for participant $\{\text{\texttt{participant id}}\}$. \\
    \cline{2-3}
    & Y3	&	Endpoint call to $\{\text{\texttt{endpoint name}}\}$ returned malformed data for participant $\{\text{\texttt{participant id}}\}$. \\
    \cline{2-3}
    & Y4 &	Endpoint call to $\{\text{\texttt{endpoint name}}\}$ returned no data for participant $\{\text{\texttt{participant id}}\}$. \\
    \cline{2-3}
    & Y5	& Treatment personalization procedure failed and fallback method was executed for participant $\{\text{\texttt{participant id}}\}$. \\
    \cline{2-3}
    & Y6	& Algorithm could not update policy (posterior parameters) with given batch data. \\
    \hline
    \caption{Oralytics monitoring issue severities. $\{\text{\texttt{endpoint name}}\}$ refers to the type of dependency endpoint that was called and $\{\text{\texttt{participant id}}\}$ refers to unique identifier assigned to the participant for the trial}
    \label{tab:oralytics_flags}
    \end{tabularx}
% \end{table*}

\section{Potential Issues for MiWaves}

% \onecolumn
% \begin{table*}[t]
    % \centering
    % \caption{MiWaves monitoring flags. EC refers to the Error Code which was used in the automated email to report for errors.}
    \begin{tabularx}{\textwidth}{ |c|c|X| }
    % \begin{xltabular}{\textwidth}{ |c|c|X| }
        % \label{tab:monitoring_miwaves} \caption{Monitoring flags for MiWaves}\\
        \hline
        Severity & EC & Description \\
        \hline
        \hline
        \endfirsthead
        \multirow{14}{*}{RED} & R1 & For a given decision point, more than 80\% of the active participants (started and did not complete 30 days yet in the trial) did not receive an intervention message.\\
        \cline{2-3}
        & R2 & For a given decision point, more than 80\% of the active users (started and did not complete 30 days yet in the trial) received an intervention message. \\
        \cline{2-3}
        & R3 & For any given user at any time, in the last 7 days, they received (i.e. get assigned)  intervention messages for more than 80\% of the decision points. \\
        \cline{2-3}
        & R4 & For any given user at any time, in the last 7 days, they did not receive (i.e. get assigned) any intervention messages for more than 80\% of the decision points. \\
        \cline{2-3}
        & R5 & The action selection probability returned by the RL algorithm for any user is greater than 0.8 or less than 0.2\\
        \hline
        \hline
        \multirow{5}{*}{YELLOW} & 0 & Authorization token is malformed.\\
        \cline{2-3}
        & 1	&	Authorization token is missing \\
        \cline{2-3}
        & 2	&	Client's authorization token is invalid/does not match \\
        \cline{2-3}
        & 3	&	Something went wrong with the authorization procedure on the RL API \\
        \cline{2-3}
        & 4	&	No more client/backends allowed to register on RL API \\
        \cline{2-3}
        & 5	&	Error occurred while trying to register client/backend on RL API \\
        \cline{2-3}
        & 6	&	Client/Backend already exists on RL API database \\
        \cline{2-3}
        & 7	&	Invalid client/backend auth credentials \\
        \cline{2-3}
        & 8	&	Something went wrong trying to authenticate client/backend credentials \\
        \cline{2-3}
        & 9	&	Failure in trying to log out client - as RL API was not able to blacklist auth token and commit to db \\
        \cline{2-3}
        & 10	&	Malformed request while trying to log out client \\
        \cline{2-3}
        & 11	&	Invalid auth token to log out\\
        \cline{2-3}
        & 100	&	User id was invalid \\
        \cline{2-3}
        & 101	&	RL start date is invalid \\
        \cline{2-3}
        & 102	&	RL end date is invalid \\
        \cline{2-3}
        & 103	&	Consent start date is invalid \\
        \cline{2-3}
        & 104	&	Consent end date is invalid \\
        \cline{2-3}
        & 105	&	Morning notification time is invalid \\
        \cline{2-3}
        & 106	&	Evening notification time is invalid \\
        \cline{2-3}
        & 107	&	Some error occurred trying to push the user registration info to RL database \\
        \cline{2-3}
        & 108	&	User has already been registered \\
        \cline{2-3}
        & 109	&	Some error occurred during user registration procedure \\
        \cline{2-3}
        & 200	&	User id was invalid \\
        \cline{2-3}
        & 201	&	Finished ema flag is invalid \\
        \cline{2-3}
        & 202	&	App use flag is invalid \\
        \cline{2-3}
        & 203	&	User id does not exist in the RL database \\
        \cline{2-3}
        & 204	&	Error trying to construct reward by the RL algorithm \\
        \cline{2-3}
        & 205	&	Error trying to construct state by the RL algorithm \\
        \cline{2-3}
        & 206	&	Error trying to compute action by the RL algorithm \\
        \cline{2-3}
        & 207	&	Some unknown error occurred trying to compute action for given user \\
        \cline{2-3}
        & 208	&	Some unknown error occurred trying to compute action for given user \\
        \cline{2-3}
        & 209	&	Activity question response field is not present when ema has been completed \\
        \cline{2-3}
        & 210	&	Cannabis use field is not present or empty when ema has been completed \\
        \cline{2-3}
        & 211	&	Window label field is not present or empty \\
        \cline{2-3}
        & 300	&	User ID does not exist \\
        \cline{2-3}
        & 301	&	The trial has not started for the given user \\
        \cline{2-3}
        & 302	&	The trial has ended for the given user \\
        \cline{2-3}
        & 303	&	Some unknown error occurred trying to commit user data and end the user decision window \\
        \cline{2-3}
        & 304	&	Invalid action taken passed by caller \\
        \cline{2-3}
        & 305	&	Invalid action selection seed passed by caller \\
        \cline{2-3}
        & 306	&	Invalid probability of action taken passed by caller \\
        \cline{2-3}
        & 307	&	Invalid policy id passed by caller \\
        \cline{2-3}
        & 308	&	Invalid decision index passed by caller \\
        \cline{2-3}
        & 309	&	Invalid action generation timestamp passed by caller \\
        \cline{2-3}
        & 310	&	Invalid action request ID (rid) passed by caller \\
        \cline{2-3}
        & 311	&	Invalid EMA finished timestamp passed by caller \\
        \cline{2-3}
        & 312	&	Invalid push notification timestamp passed by caller \\
        \cline{2-3}
        & 313	&	Invalid message click notification timestamp passed by caller \\
        \cline{2-3}
        & 314	&	Invalid morning notification time passed by caller \\
        \cline{2-3}
        & 315	&	Invalid evening notification time passed by caller \\
        \cline{2-3}
        & 316	&	Query to backend to fetch user data failed \\
        \cline{2-3}
        & 317	&	Empty data returned by backend when trying to fetch data for given user for the ending decision window \\
        \cline{2-3}
        & 318	&	More than one decision window's data returned by the backend \\
        \cline{2-3}
        & 319	&	No action found in RL database for the given action request ID (rid provided by caller) \\
        \cline{2-3}
        & 320	&	Some unknown error occurred trying to fetch data and update the decision time for given user \\
        \cline{2-3}
        & 321	&	Some error occurred while trying to update the algorithm's data records for this decision point index \\
        \cline{2-3}
        & 322	&	Decision index already exists in the database (duplicate entry returned by the backend) \\
        \cline{2-3}
        & 323	&	Some unknown error occurred trying to check if the decision index does not exist in the database \\
        \cline{2-3}
        & 324	&	Invalid window label \\
        \cline{2-3}
        & 400	&	RL API failed to dump database tables to disk \\
        \cline{2-3}
        & 401	&	RL API failed to add new RL weights to database \\
        \cline{2-3}
        & 402	&	Some unknown error occurred during RL algorithm update \\
        \cline{2-3}
        & 403	&	RL failed to update hyper-parameters \\
        \cline{2-3}
        & 404	&	RL failed to update posteriors \\
        \cline{2-3}
        & 405	&	RL update's use data flag is invoked but not implemented \\
        \cline{2-3}
        & 406	&	Error while constructing return parameters after RL update \\
        \hline
    \caption{MiWaves monitoring issue severities. EC refers to the Error Code which was used in the automated email to report for errors.}
    \label{tab:miwaves_flags}
    % \end{xltabular}
    \end{tabularx}
% \end{table*}
% \twocolumn

%% file: appendix/oralytics_database_system.tex
\section{Oralytics Database Schema}
Oralytics maintained 5 data tables:
\begin{enumerate}
    \item \textbf{Participant Info Table:} Keeps track of trial participants and participant information
    \item \textbf{Posterior Weights Table:} Stores the posterior mean and variance of the RL algorithm
    \item \textbf{Participant Data Table:} Stores every data row from every schedule formed for that participant
    \item \textbf{Treatment Selection Data Table:} Stores the participant data row corresponding to the treatment that was actually executed, including context (may or may not be imputed) used by the algorithm and the components of the reward for that participant decision point
    \item \textbf{Update Data Table:} Stores data rows corresponding to context, treatment, reward, and the treatment selection probability to update the RL algorithm
\end{enumerate}

\subsection{Data Tables and Values}

\subsubsection{Participant Info Table}

\begin{itemize}
    \item \texttt{participant\_id}: unique participant identifier
    \item \texttt{participant\_start\_day}: first day that the participant enters the trial
    \item \texttt{participant\_end\_day}: last day that the participant is in the trial
    \item \texttt{morning\_time\_weekday}: participant-specific weekday morning decision time
    \item \texttt{evening\_time\_weekday}: participant-specific weekday evening decision time
    \item \texttt{morning\_time\_weekend}: participant-specific weekend morning decision time
    \item \texttt{evening\_time\_weekend}: participant-specific weekend evening decision time
    \item \texttt{participant\_entry\_decision\_t}: trial-level decision time when the participant enters the trial
    \item \texttt{participant\_last\_decision\_t}: trial-level decision time when the participant exits the trial
    \item \texttt{currently\_in\_trial}: 1 if the participant is currently in the trial, 0 otherwise
    \item \texttt{participant\_day\_in\_trial}: participant-level day in trial [1, 70]
    \item \texttt{participant\_opened\_app}: 1 if the participant opened the app the prior day, 0 otherwise
    \item \texttt{most\_recent\_schedule\_id}: the schedule id of the schedule of actions most recently obtained by the app
\end{itemize}

\subsubsection{Posterior Weights Table}

\begin{itemize}
    \item \texttt{policy\_idx}: index for the update time and the policy, starts with 0. 0 index refers to the prior distribution before any data update and 1 index refers to the first posterior update using data.
    \item \texttt{timestamp}: timestamp of policy update
    \item \texttt{posterior\_mu.\{\}}: flattened posterior mean vector where \{\} indexes into the vector, starts with 0
    \item \texttt{posterior\_var.\{\}.\{\}}: flattened posterior covariance matrix where \{\} indexes the row and the second \{\} indexes the column, starts with 0
\end{itemize}

\subsubsection{Participant Data Table}

\begin{itemize}
    \item \texttt{participant\_id}: unique participant identifier
    \item \texttt{participant\_start\_day}: first day that the participant enters the trial
    \item \texttt{participant\_end\_day}: last day that the participant is in the trial
    \item \texttt{timestamp}: timestamp of when the action from the corresponding schedule\_id was created
    \item \texttt{schedule\_id}: id of the schedule that this action came from
    \item \texttt{participant\_decision\_t}: indexes the participant-specific decision time starts with 0, ends with 139 (note: even values denote the morning decision time, odd values denote the evening decision time)
    \item \texttt{decision\_time}: unique datetime (i.e., \%Y-\%m-\%d \%H:\%M:\%S) for when the action was executed
    \item \texttt{day\_in\_trial}: trial-level day in trial
    \item \texttt{policy\_idx}: policy used to select action
    \item \texttt{random\_seed}: random integer in [0, 999] to reproduce the Bernoulli draw of the action given action-selection probability
    \item \texttt{action int}: \{0, 1\} where 1 denotes a message being sent and 0 denotes a message not being sent
    \item \texttt{prob}: action selection probability
    \item \texttt{state.\{\}}: flattened state (context) vector observed by the algorithm at decision time (Note: this is the state used for action-selection as b\_bar, a\_bar are imputed value depending on what schedule the participant gets.)
\end{itemize}

\subsubsection{Treatment Selection Data Table}

\begin{itemize}
    \item \texttt{participant\_id}: unique participant identifier
    \item \texttt{participant\_start\_day}: first day that the participant enters the trial
    \item \texttt{participant\_end\_day}: last day that the participant is in the trial
    \item \texttt{timestamp}: timestamp of when the action from the corresponding schedule\_id was created
    \item \texttt{schedule\_id}: id of the schedule that this action came from
    \item \texttt{participant\_decision\_t}: indexes the participant-specific decision time starts with 0, ends with 139 (note: even values denote the morning decision time, odd values denote the evening decision time)
    \item \texttt{decision\_time}: unique datetime (i.e., \%Y-\%m-\%d \%H:\%M:\%S) for when the action was executed
    \item \texttt{day\_in\_trial}: trial-level day in trial
    \item \texttt{policy\_idx}: policy used to select action
    \item \texttt{random\_seed}: random integer in [0, 999] to reproduce the Bernoulli draw of the action given action-selection probability
    \item \texttt{action int}: \{0, 1\} where 1 denotes a message being sent and 0 denotes a message not being sent
    \item \texttt{prob}: action selection probability
    \item \texttt{state.\{\}}: flattened state (context) vector observed by the algorithm at decision time (Note: this is the state used for action-selection as b\_bar is an imputed value depending on what schedule the participant gets.)
    \item \texttt{brushing\_duration}: brushing durations in seconds
    \item \texttt{pressure\_duration}: pressure durations in seconds
    \item \texttt{quality}: brushing quality truncated at 180 seconds
    \item \texttt{raw\_quality}: actual brushing quality for that decision time
    \item \texttt{reward}: our designed surrogate reward given to the algorithm
    \item \texttt{cost\_term}: cost term of surrogate reward
    \item \texttt{B\_condition}: B\_condition of the cost term
    \item \texttt{A1\_condition}: A1\_condition of the cost term
    \item \texttt{A2\_condition}: A2\_condition of the cost term
    \item \texttt{actual\_b\_bar}: actual observed b\_bar state value used in calculating B\_condition (notice that a\_bar used in calculating A1\_condition and A2\_condition is the same a\_bar in the state value above)
\end{itemize}

\subsubsection{Update Data Table}

\begin{itemize}
    \item \texttt{participant\_id}: unique participant identifier
    \item \texttt{participant\_start\_day}: first day that the participant enters the trial
    \item \texttt{participant\_end\_day}: last day that the participant is in the trial
    \item \texttt{timestamp}: timestamp of when this data tuple was added to the table
    \item \texttt{participant\_decision\_t}: indexes the participant-specific decision time starts with 0, ends with 139 (note: even values denote the morning decision time, odd values denote the evening decision time)
    \item \texttt{decision\_time}: unique datetime (i.e., \%Y-\%m-\%d \%H:\%M:\%S) for when the action was executed
    \item \texttt{first\_policy\_idx}: first policy idx to use this data tuple for update
    \item \texttt{action int}: \{0, 1\} where 1 denotes a message being sent and 0 denotes a message not being sent
    \item \texttt{prob}: action selection probability
    \item \texttt{reward}: our designed surrogate reward given to the algorithm
    \item \texttt{quality}: brushing quality truncated at 180 seconds
    \item \texttt{state.\{\}}: flattened state (context) vector observed by the algorithm at decision time (Note: this is the state used to update the algorithm, b\_bar and a\_bar are the actual observed values for that decision time)
\end{itemize}

%% file: appendix/miwaves_database_system.tex
\section{MiWaves Database Schema}
MiWaves maintained 7 data tables:
\begin{enumerate}
    \item \textbf{Users Table}: Contains user-specific data such as unique identifiers, consent periods, and reinforcement learning periods (Table \ref{tab:users}).
    
    \item \textbf{User Status Table}: Tracks the status of each user within the trial, including trial phase, notification times, decision indices, and current trial day (Table \ref{tab:users_status}).
    
    \item \textbf{Algorithm Status Table}: Stores the status of the algorithm, including policy identifiers, update timestamps, and trial day information (Table \ref{tab:alg_status}).
    
    \item \textbf{RL Hyperparameter Update Request Table}: Logs requests for updating reinforcement learning hyperparameters, including request details, statuses, messages, and completion timestamps (Table \ref{tab:rl_hyperparam_update}).
    
    \item \textbf{RL Weights Table}: Contains information about the reinforcement learning model weights, including posterior means, variances, noise variance, and related metadata (Table \ref{tab:rl_weights}).
    
    \item \textbf{RL Action Selection Table}: Records the actions selected by the algorithm for each user, including decision indices, notification times, selected actions, probabilities, and rewards (Table \ref{tab:rl_action}).
    
    \item \textbf{User Action History Table}: Maintains a history of user actions and responses, including decision indices, activity responses, app usage, cannabis use data, rewards, and timestamps (Table \ref{tab:user_act_hist}).
\end{enumerate}

\begin{table*}[ht]
    \centering
    \begin{tabularx}{\textwidth}{|l|l|X|}
    \hline
    \textbf{Column Name} & \textbf{Data Type} & \textbf{Description} \\
    \hline
    user\_id & String & Unique identifier for each user \\
    consent\_start\_date & DateTime & Start date of user's consent \\
    consent\_end\_date & DateTime & End date of user's consent \\
    rl\_start\_date & DateTime & Start date of reinforcement learning period \\
    rl\_end\_date & DateTime & End date of reinforcement learning period \\
    \hline
    \caption{Users Table}
    \label{tab:users}
    \end{tabularx}
\end{table*}

\begin{table*}[t]
    \centering
    \begin{tabularx}{\textwidth}{|l|l|X|}
    \hline
    \textbf{Column Name} & \textbf{Data Type} & \textbf{Description} \\
    \hline
    user\_id & String & Foreign key referencing users table \\
    trial\_phase & Enum & Current phase of the trial \\
    morning\_notif\_time\_start & Array(Integer) & Start times for morning notifications \\
    evening\_notif\_time\_start & Array(Integer) & Start times for evening notifications \\
    current\_decision\_index & Integer & Index of the current decision point \\
    current\_time\_of\_day & Integer & Current time of day for the user \\
    trial\_day & Integer & Current day in the trial \\
    \hline
    \caption{User Status Table}
    \label{tab:users_status}
    \end{tabularx}
\end{table*}

\begin{table*}[ht]
    \centering
    % \begin{tabular}{c|c}
    %      &  \\
    %      & 
    % \end{tabular}
    % \caption{Caption}
    % \label{tab:my_label}
    \begin{tabularx}{\textwidth}{|l|l|X|}
    \hline
    \textbf{Column Name} & \textbf{Data Type} & \textbf{Description} \\
    \hline
    policy\_id & Integer & Unique identifier for the policy \\
    update\_time & DateTime & Timestamp of the last update \\
    update\_day\_in\_trial & Integer & Day in the trial when last updated \\
    current\_decision\_time & Integer & Time of the current decision \\
    current\_day\_in\_trial & Integer & Current day in the trial \\
    \hline
    \caption{Algorithm Status Table}
    \label{tab:alg_status}
    \end{tabularx}
\end{table*}

\begin{table*}[ht]
    \centering
\begin{tabularx}{\textwidth}{|l|l|X|}
\hline
\textbf{Column Name} & \textbf{Data Type} & \textbf{Description} \\
\hline
id & Integer & Unique request identifier \\
backup\_location & String & Location of the backup file \\
request\_timestamp & DateTime & Timestamp of the request \\
request\_status & String & Status of the request \\
request\_message & String & Message associated with the request \\
request\_error\_code & Integer & Error code if the request failed \\
completed\_timestamp & DateTime & Timestamp when the request was completed \\
\hline
\caption{RL Hyperparameter Update Request Table}
    \label{tab:rl_hyperparam_update}
\end{tabularx}
\end{table*}

\begin{table*}[ht]
    \centering
\begin{tabularx}{\textwidth}{|l|l|X|}
\hline
\textbf{Column Name} & \textbf{Data Type} & \textbf{Description} \\
\hline
id & Integer & Unique identifier for the RL weights record \\
policy\_id & Integer & Identifier for the policy associated with the weights \\
update\_timestamp & DateTime & Timestamp when the weights were updated \\
post\_mean\_array & Array(Float) & Array of posterior mean values \\
post\_var\_array & Array(Float) & Array of posterior variance values \\
post\_tpop\_mean\_array & Array(Float) & Array of posterior population mean values \\
post\_tpop\_var\_array & Array(Float) & Array of posterior population variance values \\
noise\_var & Float & Variance of the noise in the model \\
random\_eff\_cov\_array & Array(Float) & Covariance array for random effects \\
code\_commit\_id & String & Identifier for the code commit associated with the weights \\
data\_pickle\_file\_path & String & Path to the data pickle file \\
user\_list & Array(String) & List of users associated with the weights \\
hp\_update\_id & Integer & Identifier for the hyperparameter update request \\
\hline
\caption{RL Weights Table}
    \label{tab:rl_weights}
\end{tabularx}
\end{table*}

\begin{table*}[ht]
    \centering
\begin{tabularx}{\textwidth}{|l|l|X|}
\hline
\textbf{Column Name} & \textbf{Data Type} & \textbf{Description} \\
\hline
user\_id & String & Foreign key referencing users table \\
user\_decision\_idx & Integer & Index of the user's decision \\
morning\_notification\_time & Array(Integer) & Times for morning notifications \\
evening\_notification\_time & Array(Integer) & Times for evening notifications \\
day\_in\_trial & Integer & Current day in the trial for the user \\
action & Integer & Action selected by the algorithm \\
policy\_id & Integer & Identifier for the policy used \\
seed & Integer & Seed used for random number generation \\
prior\_ema\_completion\_time & String & Time taken to complete prior EMA \\
action\_selection\_timestamp & DateTime & Timestamp when action was selected \\
message\_sent\_notification\_ts & String & Timestamp when message was sent \\
message\_click\_notification\_ts & String & Timestamp when message was clicked \\
act\_prob & Float & Probability of the selected action \\
cannabis\_use & Array(Float) & User's cannabis use data \\
state\_vector & Array(Float) & State vector for the user \\
reward & Float & Reward received for the action \\
row\_complete & Boolean & Indicates if the row is complete \\
rid & Integer & Foreign key referencing user\_action\_history table \\
\hline
\caption{RL Action Selection Table}
    \label{tab:rl_action}
\end{tabularx}
\end{table*}

\begin{table*}[ht]
    \centering
\begin{tabularx}{\textwidth}{|l|l|X|}
\hline
\textbf{Column Name} & \textbf{Data Type} & \textbf{Description} \\
\hline
index & Integer & Unique identifier for the user action history record \\
user\_id & String & Foreign key referencing users table \\
decision\_idx & Integer & Index of the decision in the user's history \\
finished\_ema & Boolean & Indicates if the EMA was completed \\
activity\_question\_response & String & User's response to activity question \\
app\_use\_flag & Boolean & Indicates if the app was used \\
cannabis\_use & Array(Float) & User's cannabis use data \\
reward & Float & Reward associated with the action \\
state & Array(Integer) & State array for the user \\
action & Integer & Action taken by the user \\
seed & Integer & Seed used for random number generation \\
act\_prob & Float & Probability of the action taken \\
policy\_id & Integer & Identifier for the policy used \\
timestamp & DateTime & Timestamp of the action \\
\hline
\caption{User Action History Table}
    \label{tab:user_act_hist}
\end{tabularx}
\end{table*}

\section{Participant Verification Checks}
\label{app:participant_verification}
The participant verification checks were performed to verify participants for the MiWaves clinical trial, and identify any fraudulent participants. The checks involved:
\begin{enumerate}
    \item Limiting public information about eligibility criteria
    \item Using password protected screening surveys
    \item Editing screening survey settings to prevent multiple submissions from one respondent
    \item Using a CAPTCHA verification at the beginning of the screening survey to protect against bots
    \item Checking IP addresses associated with screening responses and ensure there are no duplicates
    \item Completing identity checks with eligible screens (e.g. selfie checks, ID checks, social media checks)
\end{enumerate}